\RequirePackage{ifpdf}
\ifpdf
 	\documentclass[a4paper,11pt,pdftex]{amsart}
 	\else
 	\documentclass[a4paper,11pt,dvips]{amsart}
\fi

\usepackage[all]{xy}
\ifpdf
\else
  \xyoption{ps}
  \xyoption{dvips}
\fi
\usepackage{ae,aecompl}
\usepackage[english]{babel}
\usepackage[utf8]{inputenc}
\usepackage{palatino}
\usepackage{enumerate} %For enumerating with letters etc.
\usepackage{amsthm}
\usepackage{amsmath}
\usepackage{amsfonts}
\usepackage{subfig}
\usepackage{graphicx}

\ifpdf
	\usepackage{epstopdf}		%Used to insert eps-figures in pdf files. 
	\DeclareGraphicsExtensions{.png,.jpg,.eps,.epsf}
\fi

\newcommand{\C}{\mathbf{C}}
\newcommand{\Cb}{\bar{\C}}
\newcommand{\R}{\mathbf{R}}

\newcommand{\sbraid}{\mathcal{H}}	%The Hurwitz braid group.

\newcommand{\junc}{u}		%Symbol representing a junction.
\newcommand{\vtex}{v}		%Vertex symbol.

\newcommand{\actA}{A}		%Action symbol 1
\newcommand{\actB}{B}		%Action symbol 2
\newcommand{\defin}[1]{\emph{#1}}

\mathchardef\mhyp="2D

\newtheorem{lemma}{Lemma}
\newtheorem{theorem}[lemma]{Theorem}
\newtheorem{corollary}[lemma]{Corollary}
\newtheorem{definition}[lemma]{Definition}
\newtheorem{remark}[lemma]{Remark}
\newtheorem{proposition}[lemma]{Proposition}

\title[On eigenvalues of the Schr\"odinger operator]{On eigenvalues of the Schr\"odinger operator with a complex-valued polynomial potential}

\author[P.~Alexandersson]{Per Alexandersson}
\address{Department of Mathematics, Stockholm University, SE-106 91, Stockholm, Sweden}
\email{per@math.su.se}
\author[A.~Gabrielov]{Andrei Gabrielov}
\address{Purdue University, West Lafayette, IN, 47907-2067, U.S.A.}
\email{agabriel@math.purdue.edu}

\date{\today}

\begin{document}


\begin{abstract}
In this paper, we generalize a recent result of A.~Eremenko and A.~Gabrielov on 
irreducibility of the spectral discriminant for the Schr\"odinger equation with quartic potentials. 
We consider the eigenvalue problem with a complex-valued polynomial potential
of arbitrary degree $d$ and show that the spectral determinant of this problem 
is connected and irreducible. In other words, every eigenvalue can be reached from 
any other by analytic continuation.

We also prove connectedness of the parameter spaces of the potentials that admit 
eigenfunctions satisfying $k>2$ boundary conditions, except for the case $d$ is even and $k=d/2.$ 
In the latter case, connected components of the parameter space are distinguished by the number of zeros of the eigenfunctions.
\end{abstract}

\maketitle

\tableofcontents

\section{Introduction}
\sloppy In this paper we study analytic continuation of eigenvalues of the 
Schröodinger operator with a complex-valued polynomial potential.
In other words, we are interested in the analytic continuation of eigenvalues $\lambda=\lambda(\mathbf{\alpha})$ 
of the boundary value problem for the differential equation
\begin{align}\label{eq:schroedinger}
-y'' + P_\alpha(z)y = \lambda y, 
\end{align}
where 
\begin{align*}
P_\alpha(z) = z^d + \alpha_{d-1}z^{d-1}+\dots+\alpha_1 z \mbox{ where } \mathbf{\alpha}=(\alpha_1,\alpha_2,\dots,\alpha_{d-1}), \; d\geq 2.
\end{align*}
The boundrary conditions are given by either \eqref{eq:bddconditions} or \eqref{eq:bddconditionstwo} below. 
Namely, set $n = d+2$ and divide the plane into $n$ disjoint open sectors of the form:
$$S_j = \{z \in \C \setminus \{0\} : |\arg z - 2\pi j / n|<\pi/n \}, \quad j=0,1,2,\dots,n-1.$$
These sectors are called the \defin{Stokes sectors} of the equation \eqref{eq:schroedinger}.
It is well-known that any solution $y$ of \eqref{eq:schroedinger} is either increasing or decreasing in each open Stokes sector $S_j$, i.e.
$y(z)\rightarrow 0$ or $y(z)\rightarrow \infty$ as $z \rightarrow \infty$ 
along each ray from the origin in $S_j,$ see \cite{sibuya}. 
In the first case, we say that $y$ is \defin{subdominant}, and in the second case, \defin{dominant} in $S_j.$
We impose the boundary conditions that for two \defin{non-adjacent} sectors $S_j$ and $S_k,$ i.e. for $j \neq k\pm 1\mod n:$
\begin{align}\label{eq:bddconditions}
y \text{ is } \text{ subdominant in } S_j \text{ and } S_k.
\end{align}
For example, $y(\infty)=y(-\infty)=0$ on the real axis, the boundary conditions usually imposed in physics 
for even potentials, correspond to $y$ being subdominant in $S_0$ and $S_{n/2}.$
The existence of analytic continuation is a classical fact, see e.g. references in \cite{gabrielov}.

The main results of this paper are:

\begin{theorem}\label{thm:mainone}
For any eigenvalue $\lambda_k(\alpha)$ of equation \eqref{eq:schroedinger} and boundary condition \eqref{eq:bddconditions}, there is an analytic continuation in the $\alpha\mhyp$plane to any other eigenvalue $\lambda_m(\alpha).$
\end{theorem}
We also prove some stronger results in the case where $y$ is subdominant in more than two sectors:

\begin{theorem}\label{thm:maintwo}
Given $k<n/2$ non-adjacent Stokes sectors $S_{j_1},\dots,S_{j_k},$ 
the set of all $(\alpha,\lambda) \in \C^d$ for which the equation $-y''+(P_\alpha-\lambda)y=0$ has a solution with
\begin{align}\label{eq:bddconditionstwo}
y \text{ subdominant in } S_{j_1},\dots,S_{j_k}
\end{align}
is connected.
\end{theorem}

\begin{theorem}\label{thm:mainthree}
For $n$ even and $k=n/2,$ the set of all $(\alpha,\lambda) \in \C^d$ for 
which $-y''+(P_\alpha-\lambda)y=0$ has a solution with
\begin{align}\label{eq:bddconditionsthree}
y \text{ subdominant in } S_{0},S_{2},\dots,S_{n-2}
\end{align}
is disconnected. Additionally, the solutions to \eqref{eq:schroedinger}, \eqref{eq:bddconditionstwo} have finitely many zeros, 
and the set of $\alpha$ corresponding to given number of zeros is a connected component of the former set.
\end{theorem}
The method we use is based on the ``Nevanlinna parameterization'' of 
the spectral locus introduced in \cite{gabrielov} (see also \cite{irreduc} and \cite{gabrielov10}).

\subsection{Some previous results}
In the foundational paper \cite{benderwu}, C. Bender and T. Wu studied 
analytic continuation of $\lambda$ in the complex $\beta$-plane for the problem
$$-y''+(\beta z^4+z^2)y=\lambda y, \quad y(-\infty)=y(\infty)=0.$$
Based on numerical computations, they conjectured for the first time the connectivity of the sets of odd and even eigenvalues. 
This paper generated considerable further
research in both physics and mathematics literature.
See e.g. \cite{simon} for early mathematically rigorous results in this direction.

In \cite{gabrielov}, which is the motivation of the present paper, 
the even quartic potential $P_a(z) = z^4 + a z^2$ and the boundary value problem
$$-y'' + (z^4 + a z^2)y = \lambda_a y, \quad y(\infty)=y(-\infty)=0$$
was considered. 
It is known that the problem has discrete real spectrum for real $a,$ with $\lambda_1 < \lambda_2 < \dots \rightarrow +\infty.$
There are two families of eigenvalues, those with even index and those with odd. 
The main result of \cite{gabrielov} is that if $\lambda_j$ and $\lambda_k$ are two eigenvalues in the same family, 
then $\lambda_k$ can be obtained from $\lambda_j$ by analytic continuation in the complex $\alpha\mhyp$plane.
Similar results have been obtained for other potentials, such as the PT-symmetric cubic, where $P_\alpha(z) = (iz^3+i\alpha z),$ with $y(z)\rightarrow 0,$ as $z\rightarrow \pm \infty$ on the real line. See for example \cite{irreduc}.

\begin{remark}
After this project was finished, 
the authors found out that a result similar to Theorem \ref{thm:maintwo} was proved in a hardly ever quoted Ph.D thesis, \cite{habsch}, page 36.
On the other hand, this result is formulated in the setting of Nevanlinna theory, with no connection to properties of \eqref{eq:schroedinger}.
\end{remark}

\subsection{Acknowledgements}
The second author was supported by NSF grant DMS-0801050.
Sincere thanks to Prof.~A. Eremenko for pointing out the potential relevance of \cite{habsch}.

The first author would like to thank the Mathematics department at Purdue University, for their hospitality in Spring 2010,
when this project was carried out. Also, many thanks to Boris Shapiro for being a great advisor to the first author.

%%%%%%%%%%%%%%%%%%%%%%%%%%%%%%%%%%%%%%%%%%%%%%%%%%%%%%%%%%%%%%%%%%%%%%%%%%%%%%%%%%%%%%%%%%%%%%%%%

\section{Preliminaries}
First, we recall some basic notions from Nevanlinna theory.

\begin{lemma}[see \cite{sibuya}]\label{lemma:fproperties}
Each solution $y \neq 0$ of \eqref{eq:schroedinger} is an entire function,
and the ratio $f=y/y_1$ of any two linearly independent solutions
of \eqref{eq:schroedinger} is a meromorphic function, with the 
following properties:

\begin{enumerate}[(I)]
\item For any $j,$ there is a solution $y$ of \eqref{eq:schroedinger} subdominant in the Stokes sector $S_j.$
This solution is unique, up to multiplication by a non-zero constant,

\item For any Stokes sector $S_j$, we have $f(z) \rightarrow w\in\Cb$
as $z\to\infty$ along any ray in $S_j$. This value $w$ is called
{\em the asymptotic value} of $f$ in $S_j$.

\item \label{list:different} 
For any $j$, the asymptotic values of $f$ in $S_j$ and
$S_{j+1}$ (index taken modulo $n$) are different.
The function $f$ has at least 3 distinct asymptotic values.

\item The asymptotic value of $f$ is zero in $S_j$ if and only if $y$ is subdominant in $S_j.$ 
It is convenient to call such sector \defin{subdominant} as well.
Note that the boundary conditions in \eqref{eq:bddconditions} imply that the two sectors $S_j$ and $S_k$ 
are subdominant for $f$ when $y$ is an eigenfunction of \eqref{eq:schroedinger}, \eqref{eq:bddconditions}.

\item \label{list:unramified} $f$ does not have critical points, hence
$f:\mathbf{C}\rightarrow \Cb$ is unramified outside the
asymptotic values.

\item The Schwartzian derivative $S_f$ of $f$ given by
$$S_f = \frac{f'''}{f'}-\frac{3}{2}\left( \frac{f''}{f'} \right)^2$$
equals $-2(P_\alpha - \lambda).$ Therefore one can recover $P_\alpha$ and $\lambda$ from $f$.
\end{enumerate}

\end{lemma}
From now on, $f$ always denotes the ratio of two linearly independent solutions of \eqref{eq:schroedinger}, 
with $y$ being an eigenfunction of the boundary value problem \eqref{eq:schroedinger}, 
with conditions \eqref{eq:bddconditions}, \eqref{eq:bddconditionstwo} or \eqref{eq:bddconditionsthree}.

\subsection{Cell decompositions}
Set $n = d + 2,$ $d=\deg P$ where $P$ is our polynomial potential and assume that all non-zero asymptotic values of $f$ are distinct and finite.
Let $w_j$ be the asymptotic values of $f,$ ordered arbitrarily with the only restriction that $w_j=0$ if and only if $S_j$ is subdominant.
For example, one can denote by $w_j$ the asymptotic value in the Stokes sector $S_j.$
We will later need different orders of the non-zero asymptotic values, see section \ref{subsec:standardorder}.

\begin{figure}
 \centering
   \subfloat[(a) $\Psi_0$]{\includegraphics[width=0.48\textwidth]{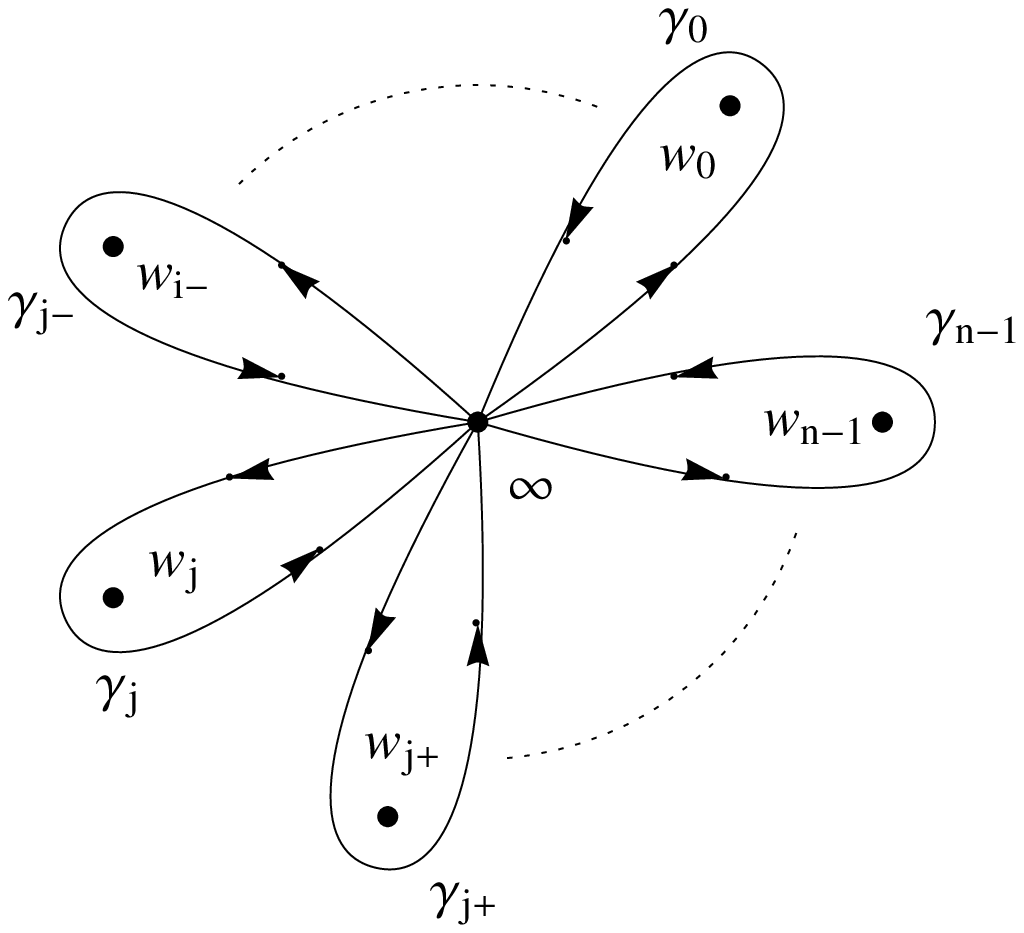}}
   \subfloat[(b) $\actA_j(\Psi_0).$]{\includegraphics[width=0.48\textwidth]{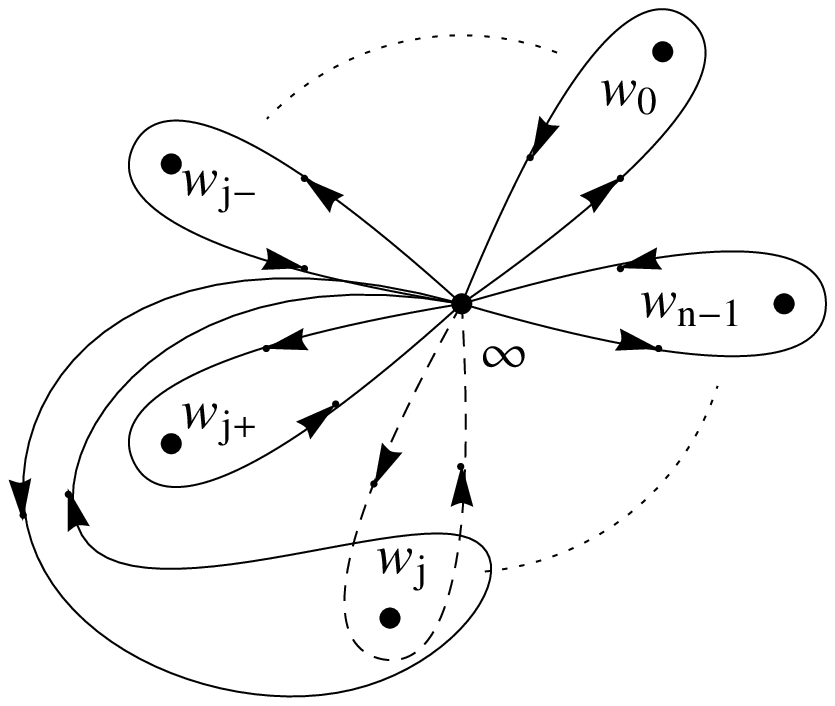}}
   \caption{Permuting $w_j$ and $w_{j_+}$ in $\Psi_0.$}\label{fig:curves}
\end{figure}

Consider the cell decomposition $\Psi_0$ of $\Cb_w$ shown in Fig.~\ref{fig:curves}a.
It consists of closed directed loops $\gamma_j$ starting and ending at $\infty,$ where the index is considered mod $n,$ and $\gamma_j$ is defined only if $w_j\neq 0.$ The loops $\gamma_j$ only intersect at $\infty$ and have no self-intersection other than $\infty.$
Each loop $\gamma_j$ contains a single non-zero asymptotic value $w_j$ of $f.$
For example, the boundary condition $y\rightarrow 0$ as $z\rightarrow \pm\infty$ for $z \in \R$ for even $n$ implies that $w_0=w_{n/2}=0,$
so there are no loops $\gamma_0$ and $\gamma_{n/2}.$ 
We have a natural cyclic order of the asymptotic values, namely the order in which a small circle around $\infty$ counterclockwise intersects the associated loops $\gamma_j,$ see Fig.~\ref{fig:curves}a.

We use the same index for the asymptotic values and the loops, which motivates the following notation:
$$j_+ = j+k \text{ where } k\in\{1,2\} \text{ is the smallest integer such that } w_{j+k}\neq 0.$$
Thus, $\gamma_{j_+}$ is the loop around the next to $w_j$ (in the cyclic order mod $n$) non-zero asymptotic value. Similarly, $\gamma_{j_-}$ is the loop around the previous non-zero asymptotic value.

\subsection{From cell decompositions to graphs}\label{sec:graphprop}

\begin{figure}
\centering
  \subfloat[$\Gamma$]{\includegraphics[width=0.33\textwidth]{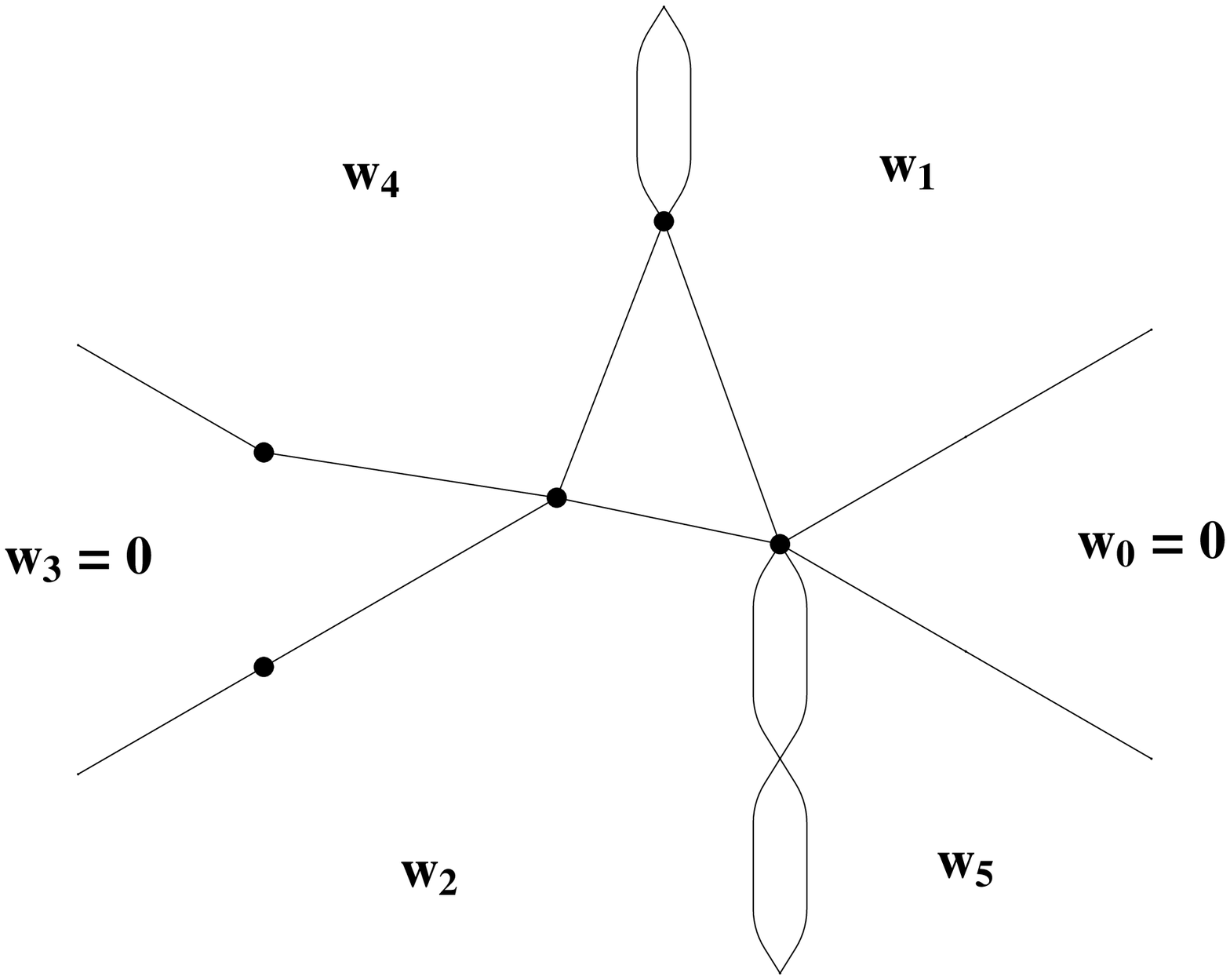}}
  \hfill
  \subfloat[$T_\Gamma$]{\includegraphics[width=0.33\textwidth]{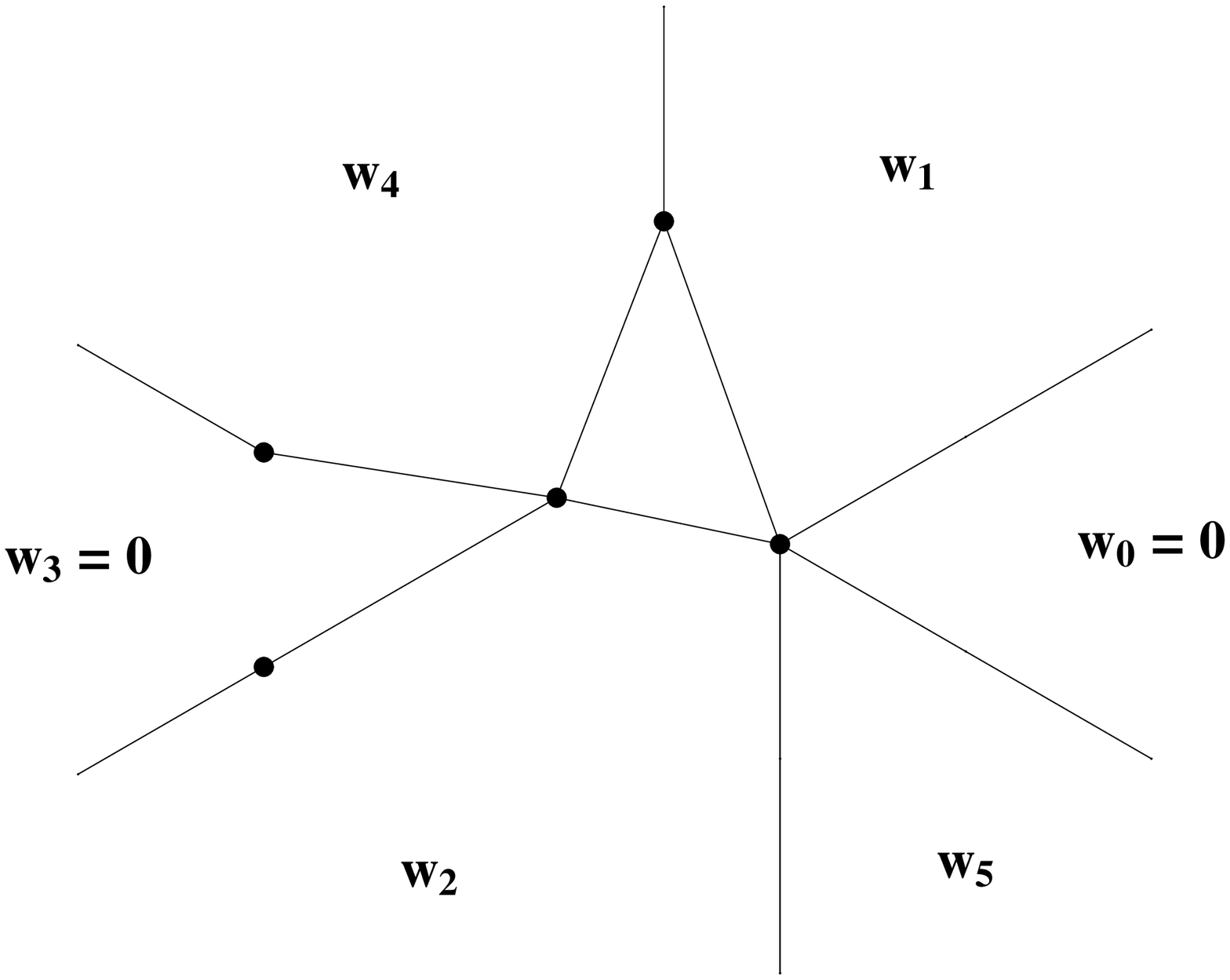}}
  \hfill
  \subfloat[$T^*_\Gamma$]{\includegraphics[width=0.33\textwidth]{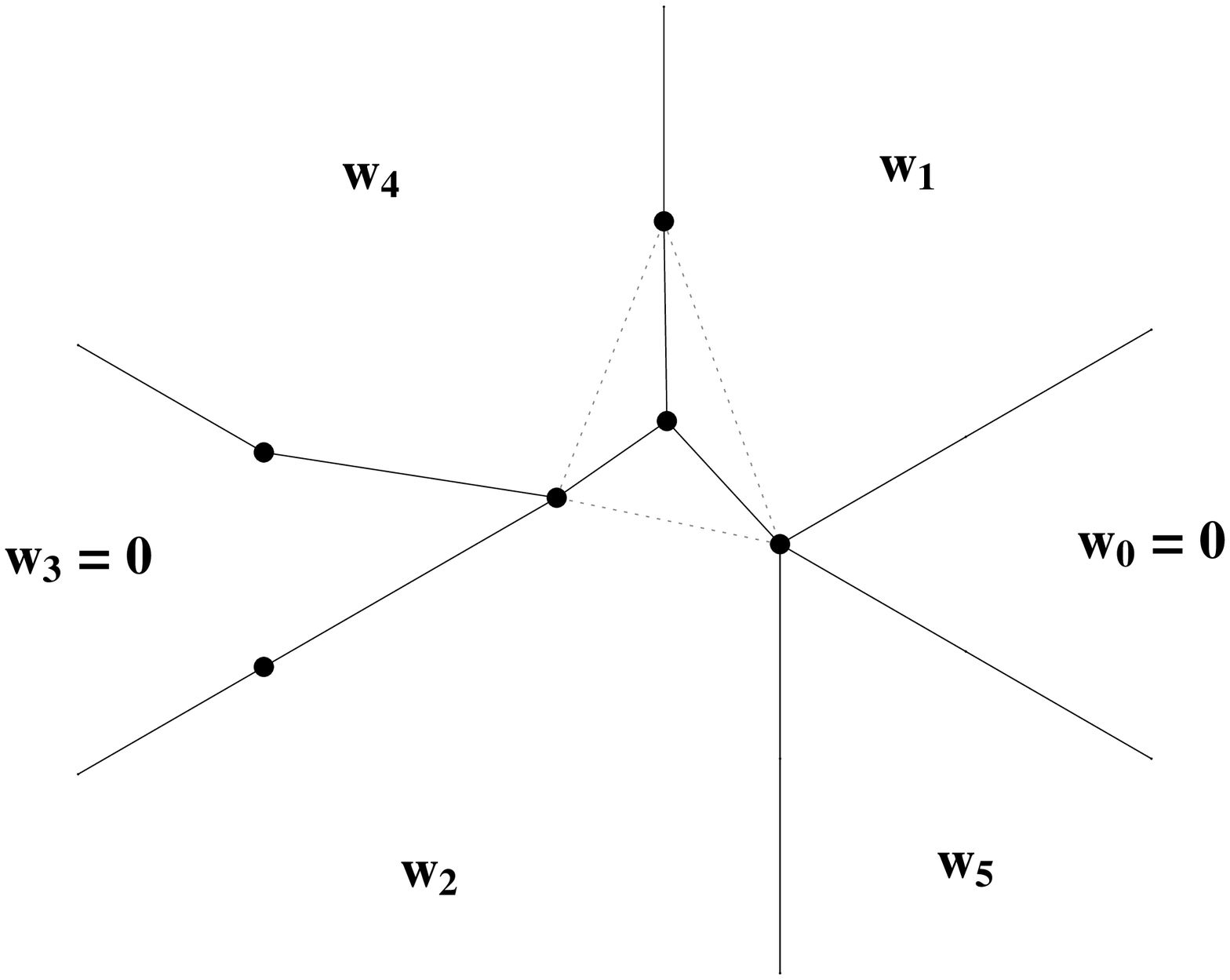}}
  \caption{The correspondence between $\Gamma,$ $T_\Gamma$ and $T^*_\Gamma$.}
  \label{fig:example_cell_complex}
\end{figure}

We may simplify our work with cell decompositions with the help of the following:

\begin{lemma}[See Section 3 \cite{gabrielov}]
Given $\Psi_0$ as in Fig.~\ref{fig:curves}a, one has the following properties:

\begin{enumerate}[(a)]
\item The preimage $\Phi_0 = f^{-1}(\Psi_0)$ gives a cell decomposition of the plane $\C_z.$ 
Its vertices are the poles of $f,$ and the edges are preimages of the loops $\gamma_j.$ 
These edges are labeled by $j,$ and are called $j\mhyp$edges.

\item The edges of $\Phi_0$ are directed, their orientation is induced from the orientation of the loops $\gamma_j$.  
Removing all loops of $\Phi_0,$ we obtain an infinite, directed planar graph $\Gamma,$ without loops. 

\item Vertices of $\Gamma$ are poles of $f,$ each bounded connected component of $\C\setminus \Gamma$ 
contains one simple zero of $f,$ and each zero of $f$ belongs to one such bounded connected component.

\item There are at most two edges of $\Gamma$ connecting any two of its vertices. 
Replacing each such pair of edges with a single undirected edge and making all other edges undirected, 
we obtain an undirected graph $T_\Gamma.$ 

\item $T_\Gamma$ has no loops or multiple edges, and the transformation from $\Phi_0$ to $T_\Gamma$ can be uniquely reversed.
\end{enumerate}
\end{lemma}
An example of the transformation from $\Gamma$ to $T_\Gamma$ is presented in Fig.~\ref{fig:example_cell_complex}.

A \defin{junction} is a vertex of $\Gamma$ (and of $T_\Gamma$) at which the degree of $T_\Gamma$ is at least 3.
From now on, $\Gamma$ refers to both the directed graph without loops and the associated cell decomposition $\Phi_0$.

\subsection{The standard order}\label{subsec:standardorder}
For a potential of degree $d,$ the graph $\Gamma$ has $d+2=n$ infinite branches and $n$ unbounded faces corresponding to the Stokes sectors.
We defined earlier the ordering $w_0,w_1,\dots,w_{n-1}$ of the asymptotic values of $f.$ 

\emph{If} each $w_j$ is the asymptotic value in the sector $S_{j},$ we say that the asymptotic values have \defin{the standard order} and the corresponding cell decomposition $\Gamma$ is a \defin{standard graph}.

\begin{lemma}[See Prop 6. \cite{gabrielov}]
If a cell decomposition $\Gamma$ is a standard graph, the corresponding undirected graph $T_\Gamma$ is a tree.
\end{lemma}
This property is essential in the present paper, and we classify cell decompositions of this type by describing the associated trees.

Below we define the action of the braid group that permute non-zero asymptotic values of $\Psi_0.$ 
This induces the corresponding action on graphs.
Each unbounded face of $\Gamma$ (and $T_\Gamma$) will be labeled by the asymptotic value in the corresponding Stokes sector. 
For example, labeling an unbounded face corresponding to $S_k$ with $w_j$ or just with the index $j,$ 
we indicate that $w_j$ is the asymptotic value in $S_k.$

From the definition of the loops $\gamma_j,$ a face corresponding to a dominant sector has the same label as any edge bounding that face. 
The label in a face corresponding to a subdominant sector $S_k$ is always $k,$ since the actions defined below only permute non-zero asymptotic values.
We say that an unbounded face of $\Gamma$ is (sub)dominant if the corresponding Stokes sector is (sub)dominant.

For example, in Fig.~\ref{fig:example_cell_complex}, the Stokes sectors $S_0$ and $S_3$ are subdominant since the corresponding faces have label $0.$ 
We do not have the standard order for $\Gamma,$ since $w_2$ is the asymptotic value for $S_4,$ and $w_4$ is the asymptotic value for $S_2.$ The associated graph $T_\Gamma$ is not a tree.

\subsection{Properties of graphs and their face labeling}\label{subsec:labelingprop}\label{subsec:junctionoftype}
\begin{lemma}[see \cite{gabrielov}]\label{lemma:labelingprop}
%TODO: page ref in article?
The following holds:

\begin{enumerate}[(I)]
\item Two bounded faces of $\Gamma$ cannot have a common edge, since a $j\mhyp$edge is always at the boundary of an unbounded face labeled $j.$

\item \label{list:numbering} The edges of a bounded face of a graph $\Gamma$ are directed clockwise, and their labels increase in that order. Therefore,
a bounded face of $T_\Gamma$ can only appear if the order of $w_j$ is non-standard. 

(As an example, the bounded face in Fig.~\ref{fig:example_cell_complex} has the labels $1, 2, 4$ (clockwise) of its boundary edges.)

\item \label{list:uniquenr} Each label appears at most once in the boundary of any bounded face of $\Gamma.$

\item Unbounded faces of $\Gamma$ adjacent to its junction $\junc$
always have the labels cyclically increasing counterclockwise around $\junc.$

\item \label{list:representation} To each graph $T_\Gamma,$ we associate a tree by inserting a new vertex inside each of its bounded faces, 
connecting it to the vertices of the bounded face and removing the boundrary edges of the original face. 
Thus we may associate a tree $T^*_\Gamma$ with \emph{any} cell decomposition, not necessarily with standard order, as in Fig.~\ref{fig:example_cell_complex}(c).
The order of $w_j$ above together with this tree uniquely determines $\Gamma.$ This is done using the two properties above.

\item \label{list:labelexistone} The boundary of a dominant face labeled $j$ consists of infinitely many directed $j\mhyp$edges,
oriented counterclockwise around the face.

\item \label{list:labelexisttwo} If $w_j = 0$ there are no $j\mhyp$edges.

\item Each vertex of $\Gamma$ has even degree, since each vertex in $\Phi_0 = f^{-1}(\Psi_0)$ has even degree, 
and removing loops to obtain $\Gamma$ preserves this property.
\end{enumerate}
\end{lemma}

Following the direction of the $j\mhyp$edges, the first vertex that is connected to an edge labeled ${j_+}$ is the vertex where the $j\mhyp$edges and the ${j_+}\mhyp$edges \defin{meet}. The last such vertex is where they \defin{separate}. These vertices, if they exist, must be junctions.
\begin{definition}\label{def:jjunction}
Let $\Gamma$ be a standard graph, and let $j \in \Gamma$ be a junction where the $j\mhyp$edges and $j_+\mhyp$edges separate.
Such junction is called a \defin{$j\mhyp$junction.}
\end{definition}
There can be at most one $j\mhyp$junction in $\Gamma,$ the existence of two or more such junctions would violate property (\ref{list:uniquenr}) of the face labeling. 
However, the same junction can be a $j\mhyp$junction for different values of $j.$

\begin{figure}[ht!]
\centering
  \subfloat[(a) $I\mhyp$structure.]{\includegraphics{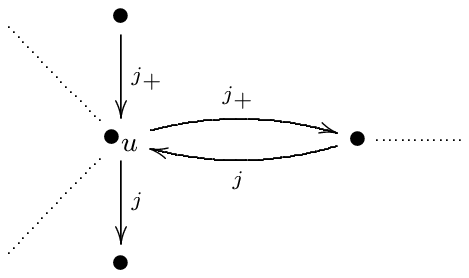}}
  \hfill
  \subfloat[(b) $V\mhyp$structure.]{\includegraphics{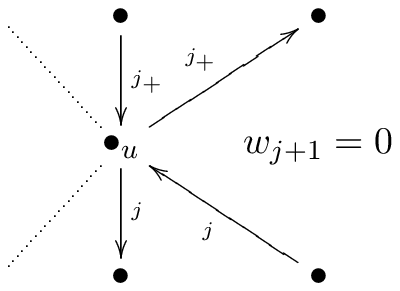}}
  \hfill
  \subfloat[(c) $Y\mhyp$structure.]{\includegraphics{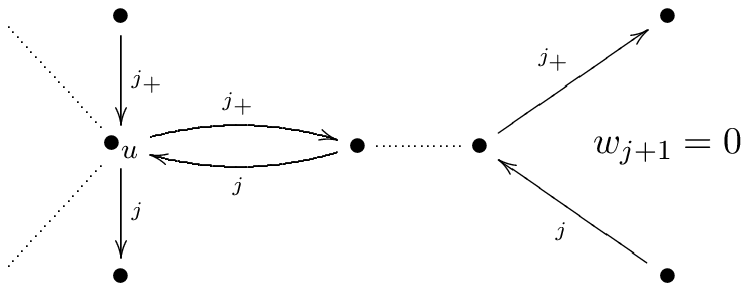}}
  \caption{Different types of $j\mhyp$junctions.}
  \label{fig:junctiontype_case123}
\end{figure}

There are three different types of $j\mhyp$junctions, see Fig.~\ref{fig:junctiontype_case123}. 

Case (a) only appears when $w_{j+1}\neq 0.$ 
Cases (b) and (c) can only appear when $w_{j+1}=0.$ In (c), the $j\mhyp$edges and $j_+\mhyp$edges meet and separate at different junctions,
while in (b), this happens at the same junction.

\begin{definition}
Let $\Gamma$ be a standard graph with a $j\mhyp$junction $u$.
A \emph{structure} at the $j\mhyp$junction is the subgraph $\Xi$
of $\Gamma$ consisting of the following elements:
\begin{itemize}
\item The edges labeled $j$ that appear before $\junc$ following the $j\mhyp$edges.
\item The edges labeled $j_+$ that appear after $\junc$ following the $j_+\mhyp$edges.
\item All vertices the above edges are connected to.
\end{itemize}
If $\junc$ is as in Fig.~\ref{fig:junctiontype_case123}a, $\Xi$ is called an \defin{$I\mhyp$structure at the $j\mhyp$junction}.
If $\junc$ is as in Fig.~\ref{fig:junctiontype_case123}b,
$\Xi$ is called a \defin{$V\mhyp$structure at the $j\mhyp$junction}.
If $\junc$ is as in Fig.~\ref{fig:junctiontype_case123}c,
$\Xi$ is called a \defin{$Y\mhyp$structure at the $j\mhyp$junction}.
\end{definition}
Since there can be at most one $j\mhyp$junction, there can be at most one structure at the $j\mhyp$junction.

\begin{figure}
\centering
\includegraphics[width=0.6\textwidth]{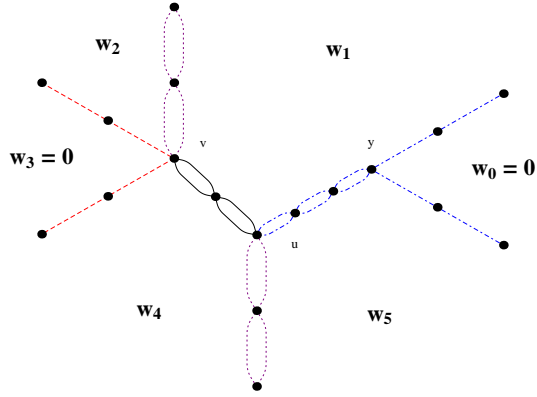}
\caption{Graph $\Gamma$ with (dotted) $I\mhyp$structures, a (dashed) $Y\mhyp$structure and a (dotdashed) $Y\mhyp$structure.}\label{fig:example_structs}
\end{figure}

A graph $\Gamma$ shown in Fig.~\ref{fig:example_structs} has one (dotted) $I\mhyp$structure at the $1\mhyp$junction $v,$
one (dotted) $I\mhyp$structure at the $4\mhyp$junction $u,$
one (dashed) $V\mhyp$structure at the $2\mhyp$junction $v$ and one (dotdashed) $Y\mhyp$structure at the $5\mhyp$junction $u$.

Note that the $Y\mhyp$structure is the only kind of structure that contains an additional junction.
We refer to such junctions as \defin{$Y\mhyp$junctions}. For example, the junction marked $y$ in Fig.~\ref{fig:example_structs} is a $Y\mhyp$junction.

\subsection{Describing trees and junctions}

Let $\Gamma$ be a graph with $n$ branches, and $\Lambda$ be the associated tree with all non-junction vertices removed.
The dual graph $\hat{\Lambda}$ of $\Lambda,$ is an $n\mhyp$gon where some non-intersecting chords are present.
The junctions of $\Lambda$ is in one-to-one correspondence with faces of $\hat{\Lambda}$ and vice versa.
Two vertices are connected with an edge in $\hat{\Lambda}$ if and only if the corresponding faces are adjacent in $\Lambda.$

The extra condition that subdominant faces do not share an edge, implies that
there are no chords connecting vertices in $\hat{\Lambda}$ corresponding to subdominant faces.
For trees without this condition, we have the following lemma:
\begin{lemma}
The number of $n+1\mhyp$gons with non-intersecting chords is equal to the number of bracketings of 
a string with $n$ letters, such that each bracket pair contains at least two symbols.
\end{lemma}
\begin{proof}
See Theorem 1 in \cite{schroedernumbers}. 
%The bijection is given as follows:Let the word with $n$ symbols be $1,2,3,\dots,n$ and let the vertices in the $n+1\mhyp$gon be labelled $0,1,2,\dots,n$ clockwise
\end{proof}
The sequence $s(n)$ of bracketings of a string with $n+1$ symbols are called the small Schröder numbers,
see \cite{schroedernumbers}. The first entries are $s(n)_{n\geq 0} = 1,1,3,11,45,197,\dots.$

The condition that chords should not connect vertices corresponding to subdominant faces,
translates into a condition on the first and last symbol in some bracket pair.

% Hence, describing the chords in $\hat{\Lambda}$ and the number of vertices between the junctions in $T_\Gamma$
% uniquely specifies $\Gamma.$ As an example,
% $$(0,1,3,5),(1,2,3),(3,4,5)$$
% describes the faces in a hexagon with the vertices enumerated counter-clockwise 0 to 5.
% The chords are $(0,1),(1,3),(3,5),(1,2),(2,3),(3,4),(4,5).$

%%%%%%%%%%%%%%%%%%%%%%%%%%%%%%%%%%%%%%%%%%%%%%%%%%%%%%%%%%%%%%%%%%%%%%%%%%%%%%%%%%%%%%%%%%%%%%%%%%%%%%%%%%%%%%%%%%%%%%%%%

\section{Actions on graphs}\label{subsec:actiondef}
\subsection{Definitions}
Let us now return to the cell decomposition $\Psi_0$ in Fig.~\ref{fig:curves}a.
Let $w_j$ be a non-zero asymptotic value of $f$. Choose non-intersecting paths $\beta_j(t)$ and $\beta_{j_+}(t)$ in $\Cb_w$ with 
$\beta_j(0) = w_j, \; \beta_j(1)=w_{j_+}$ and $\beta_{j_+}(0) = w_{j_+},$ $\beta_{j_+}(1)=w_{j}$ so that they do not intersect
$\gamma_k$ for $k\neq j,j_+$ and such that the union of these paths is a simple contractible loop oriented counterclockwise.
These paths define a continuous deformation of the loops $\gamma_j$ and $\gamma_{j_+}$ such that the two deformed loops contain
$\beta_j(t)$ and $\beta_{j_+}(t),$ respectively, 
and do not intersect any other loops during the deformation (except at $\infty$).
We denote the action on $\Psi_0$ given by $\beta_j(t)$ and $\beta_{j_+}(t)$ by $\actA_j.$
Basic properties of the fundamental group of a punctured plane, 
allows one to express the new loops in terms of the old ones:
$$
\actA_j(\gamma_k) =
\begin{cases}
\gamma_j\gamma_{j_+}\gamma_j^{-1} \text{ if } k=j\\
\gamma_j\text{ if } k=j_+ \\
\gamma_k \text{ otherwise }
\end{cases},
\quad
\actA_j^{-1}(\gamma_k) =
\begin{cases}
\gamma_{j_+} \text{ if } k=j \\
\gamma_{j_+}^{-1}\gamma_{j}\gamma_{j_+} \text{ if } k=j_+\\
\gamma_k \text{ otherwise }
\end{cases}
$$
Let $f_t$ be a deformation of $f$.
Since a continuous deformation does not change the
graph, the deformed graph
corresponding to $f_1^{-1}(\actA_j(\Psi_0))$
is the same as $\Gamma$.
Let $\Gamma'$ be this deformed graph
with labels $j$ and $j_+$ exchanged.
Then the $j\mhyp$edges of $\Gamma'$ are
$f_1^{-1}(A_j(\gamma_{j_+}))=f_1^{-1}(\gamma_j)$,
hence they are the same as the $j\mhyp$edges of $A_j(\Gamma)$.
The $j_+\mhyp$edges of $\Gamma'$ are $f_1^{-1}(A_j(\gamma_j))$.
Since $\gamma_{j_+}=\gamma_j^{-1}A_j(\gamma_j)\gamma_j,$
(reading left to right)
this means that a $j_+\mhyp$edge of $A_j(\Gamma)$ is obtained
by moving backwards along a $j\mhyp$edge of $\Gamma'$,
then along a $j_+$-edge of $\Gamma'$,
followed by a $j$-edge of $\Gamma'$.

These actions, together with their inverses, generate the Hurwitz (or sphere) braid group $\sbraid_m,$ where $m$ is the number of non-zero asymptotic values.
For a definition of this group, see \cite{lando}. 
The action $\actA_j$ on the loops in $\Psi_0$ is presented in Fig.~\ref{fig:curves}b. 

The property (\ref{list:unramified}) of the eigenfunctions implies that each $\actA_j$ induces a monodromy transformation of the cell decomposition $\Phi_0,$ and of the associated directed graph $\Gamma.$

Reading the action \textit{right to left} gives the new edges in terms of the old ones, as follows:

Applying $\actA_j$ to $\Gamma$ can be realized by first interchanging the labels $j$ and $j_+.$ This gives an intermediate graph $\Gamma'.$
A $j$-edge of $\actA_j(\Gamma)$ starting at the vertex $v$ ends at a vertex
obtained by moving from $v$ following first the $j$-edge of $\Gamma'$
backwards, then the $j_+$-edge of $\Gamma'$, and finally the
$j$-edge of $\Gamma'$. 
If any of these edges does not exist, we just do not move. If we end up at the same vertex $v$,
there is no $j$-edge of $A_j(\Gamma)$ starting at $v$. All $k$-edges of $A_j(\Gamma)$ for $k \neq j$ are the same as $k$-edges of $\Gamma'.$

\begin{figure}
\centering
  \subfloat[$\Gamma$]{\includegraphics[width=0.33\textwidth]{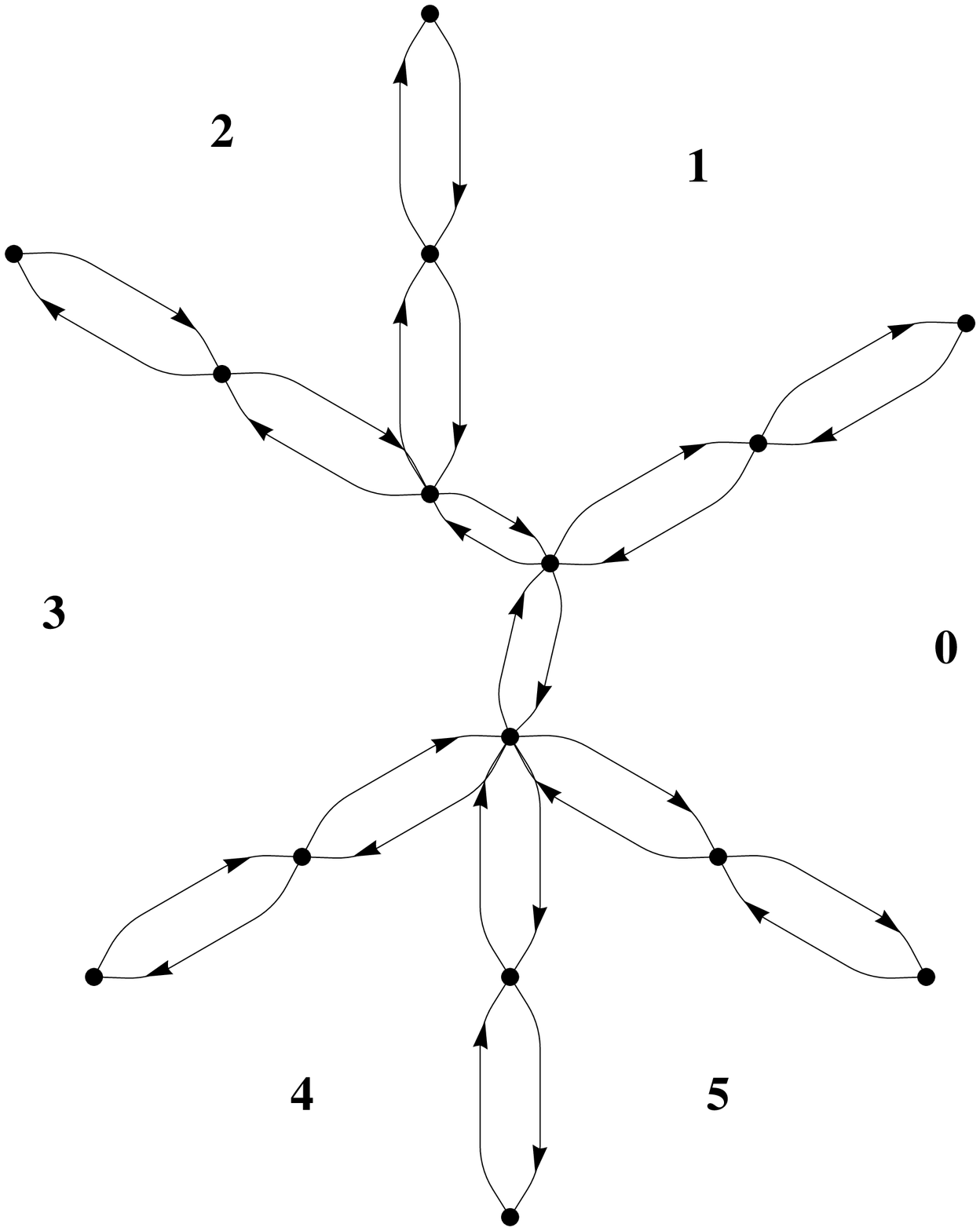}}
  \hfill
  \subfloat[$\Gamma'$]{\includegraphics[width=0.33\textwidth]{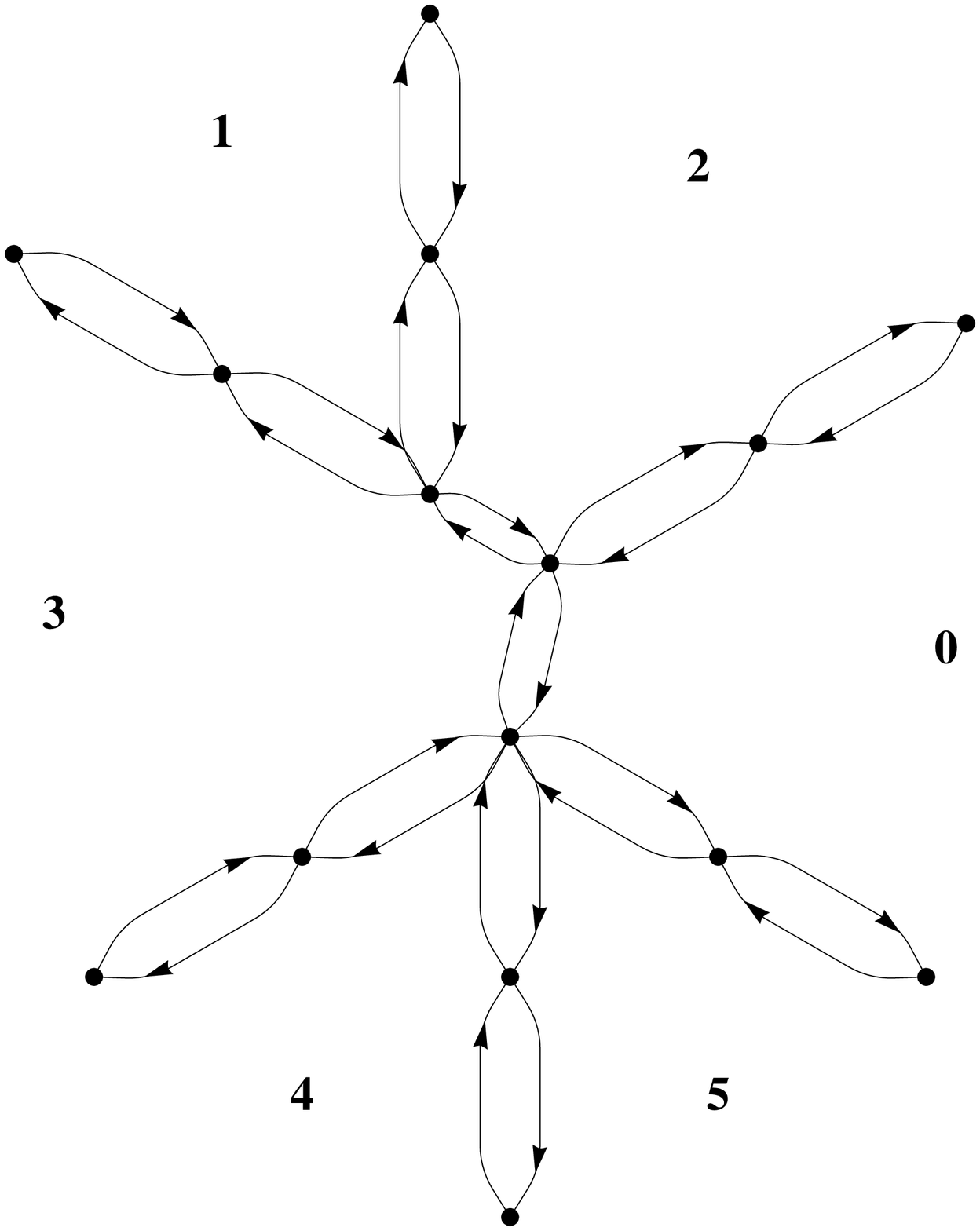}}
  \hfill
  \subfloat[$\actA_1(\Gamma)$]{\includegraphics[width=0.33\textwidth]{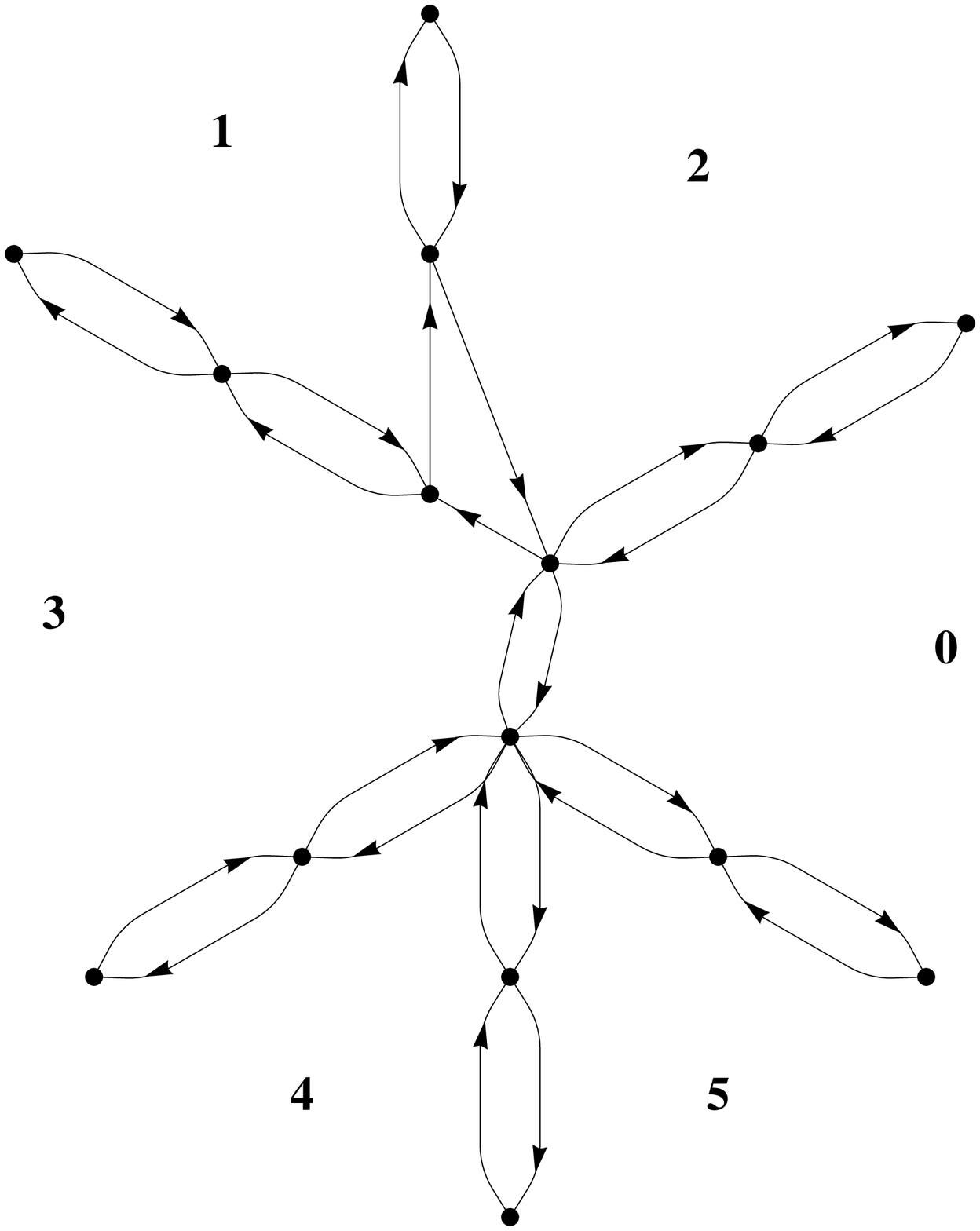}}
  \caption{The action $\actA_1.$ All sectors are dominant.}
  \label{fig:slaex}
\end{figure}

An example of the action $\actA_1$ is presented in Fig.~\ref{fig:slaex}. 
Note that $A_j^2$ preserves the standard cyclic order. 

\subsection{Properties of the actions}\label{sec:action}

\begin{lemma}\label{lemma:noaction}
Let $\Gamma$ be a standard graph with no $j\mhyp$junction. Then $\actA_j^2(\Gamma)=\Gamma.$
\end{lemma}
\begin{proof}
Since we assume $d>2$, lemma \ref{lemma:labelingprop} implies that
the boundaries of the faces of $\Gamma$ labeled $j$ and $j_+$
do not have a common vertex.
From the definition of the actions in subsection \ref{subsec:actiondef}, 
the graphs $\Gamma$ and $\actA_j(\Gamma)$ are the same,
except that the labels $j$ and $j_+$ are permuted.
Applying the same argument again gives $\actA_j^2(\Gamma)=\Gamma.$
\end{proof}

\begin{figure}[ht!]
\centering
\includegraphics[width=0.99\textwidth]{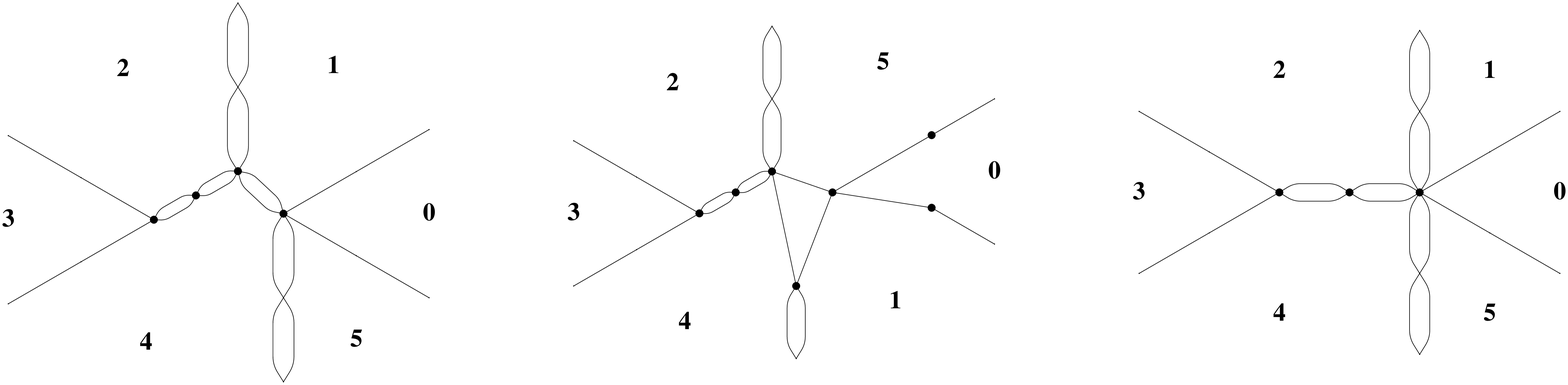}
\caption{Example action of $\actA_5^{-1}$ and $\actA_5^{-2}$ in case 3.}\label{fig:action_example_case2}
\end{figure}

\begin{theorem}\label{thm:action}
Let $\Gamma$ be a standard graph with a $j\mhyp$junction $\junc.$
Then $\actA_j^2(\Gamma)\neq \Gamma,$ 
and the structure at the $j\mhyp$junction
is moved one step in the direction of the $j\mhyp$edges under $\actA_j^2.$
The inverse of $\actA^2_j$ moves the structure at the $j\mhyp$junction one step backwards along the $j_+\mhyp$edges.
\end{theorem}
\begin{proof}
There are three cases to consider, namely $I\mhyp$structures, $V\mhyp$structures and $Y\mhyp$structures resp.

\noindent\textbf{Case 1:} The structure at the $j\mhyp$junction is an $I\mhyp$structure and 
$\Gamma$ is as in Fig.~\ref{fig:si_action_case1}a.
The action $\actA_j$ first permutes the asymptotic values $w_j$ and $w_{j_+},$ 
then transforms the new $j\mhyp$ and $j_+\mhyp$edges, as defined in subsection \ref{subsec:actiondef}.
The resulting graph $\actA_j(\Gamma)$ is shown in Fig.~\ref{fig:si_action_case1}b.
Applying $\actA_j$ to $\actA_j(\Gamma)$ yields the graph shown in Fig.~\ref{fig:si_action_case1}c. 

\begin{figure}
\centering
  \subfloat[(a) Graph $\Gamma$ with an $I$-structure]{\includegraphics[width=0.8\textwidth]{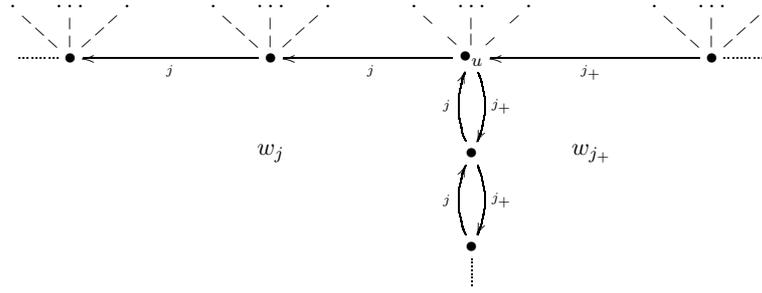}}\\
  \subfloat[(b) Graph $A_j(\Gamma)$]{\includegraphics[width=0.8\textwidth]{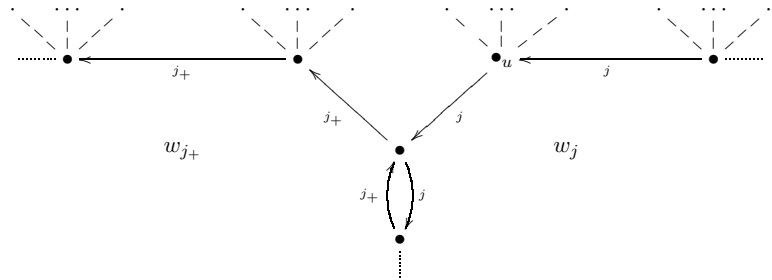}}\\
  \subfloat[(c) Graph $A_j^2(\Gamma)$]{\includegraphics[width=0.8\textwidth]{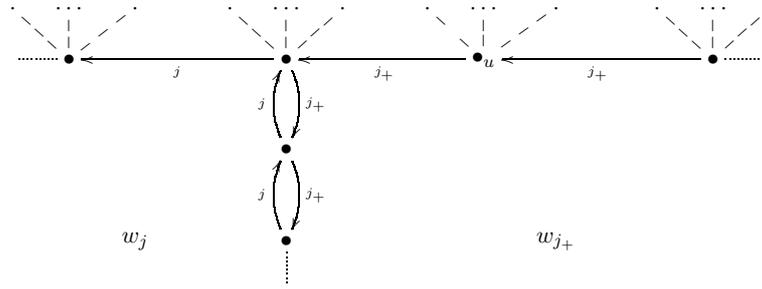}}
  \caption{Case 1, moving an $I\mhyp$structure.}
  \label{fig:si_action_case1}
\end{figure}

\noindent\textbf{Case 2:} The structure at the $j\mhyp$junction is a $V\mhyp$structure and $\Gamma$ is as in Fig.~\ref{fig:si_action_case3}a.
The graphs $\actA_j(\Gamma)$ and $\actA_j^2(\Gamma)$ are as in Fig.~\ref{fig:si_action_case3}b and in Fig.~\ref{fig:si_action_case3}c respectively.
\begin{figure}
\centering
  \subfloat[(a) Graph $\Gamma$ with a $V\mhyp$structure]{\includegraphics[width=0.8\textwidth]{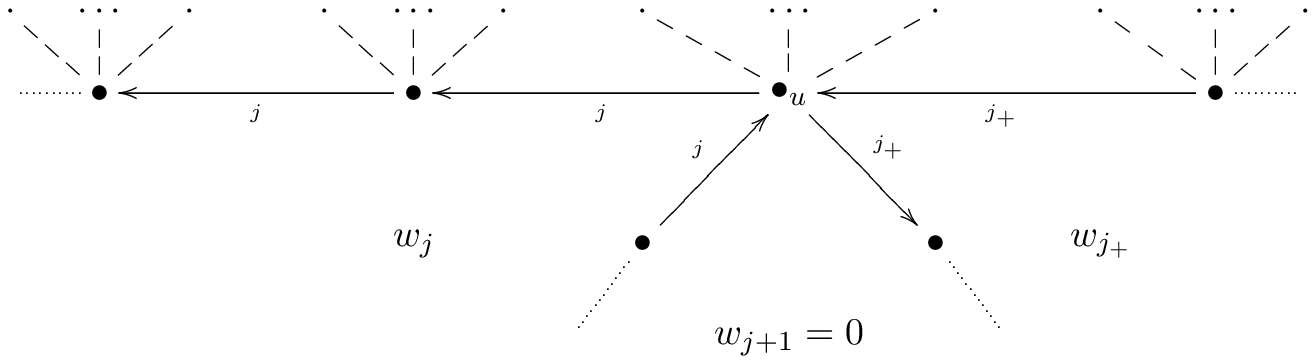}}\\
  \subfloat[(b) Graph $A_j(\Gamma)$]{\includegraphics[width=0.8\textwidth]{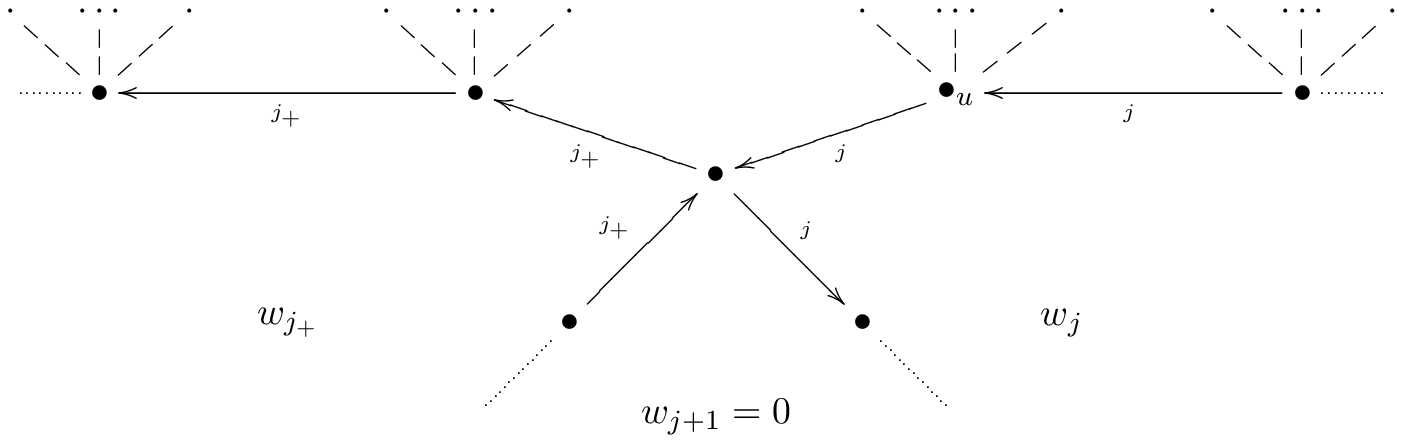}}\\
  \subfloat[(c) Graph $A_j^2(\Gamma)$]{\includegraphics[width=0.8\textwidth]{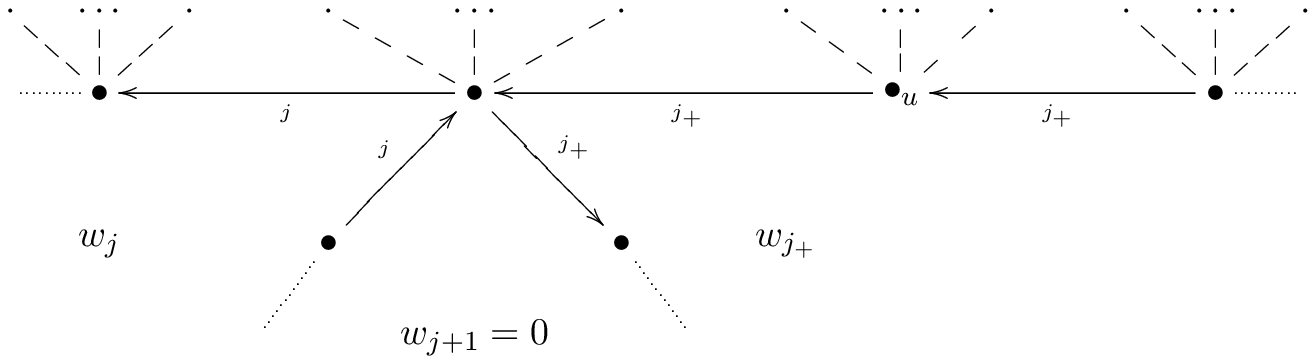}}
  \caption{Case 2, moving a $V\mhyp$structure.}
  \label{fig:si_action_case3}
\end{figure}

\noindent\textbf{Case 3:} The structure at the $j\mhyp$junction is a $Y\mhyp$structure and $\Gamma$ is as in Fig.~\ref{fig:si_action_case2}a.
The graphs $\actA_j(\Gamma)$ and $\actA_j^2(\Gamma)$ 
are as in Fig.~\ref{fig:si_action_case2}b and in Fig.~\ref{fig:si_action_case2}c respectively.
\begin{figure}
\centering
  \subfloat[(a) Graph $\Gamma$ with a $Y\mhyp$structure]{\includegraphics[width=0.8\textwidth]{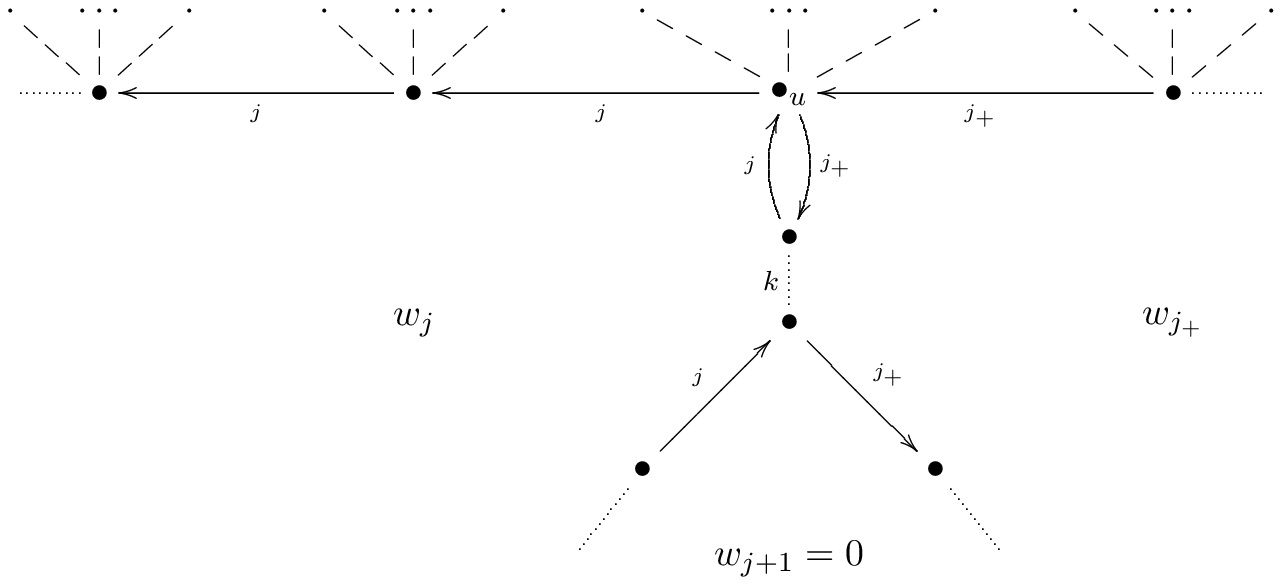}}\\
  \subfloat[(b) Graph $A_j(\Gamma)$]{\includegraphics[width=0.8\textwidth]{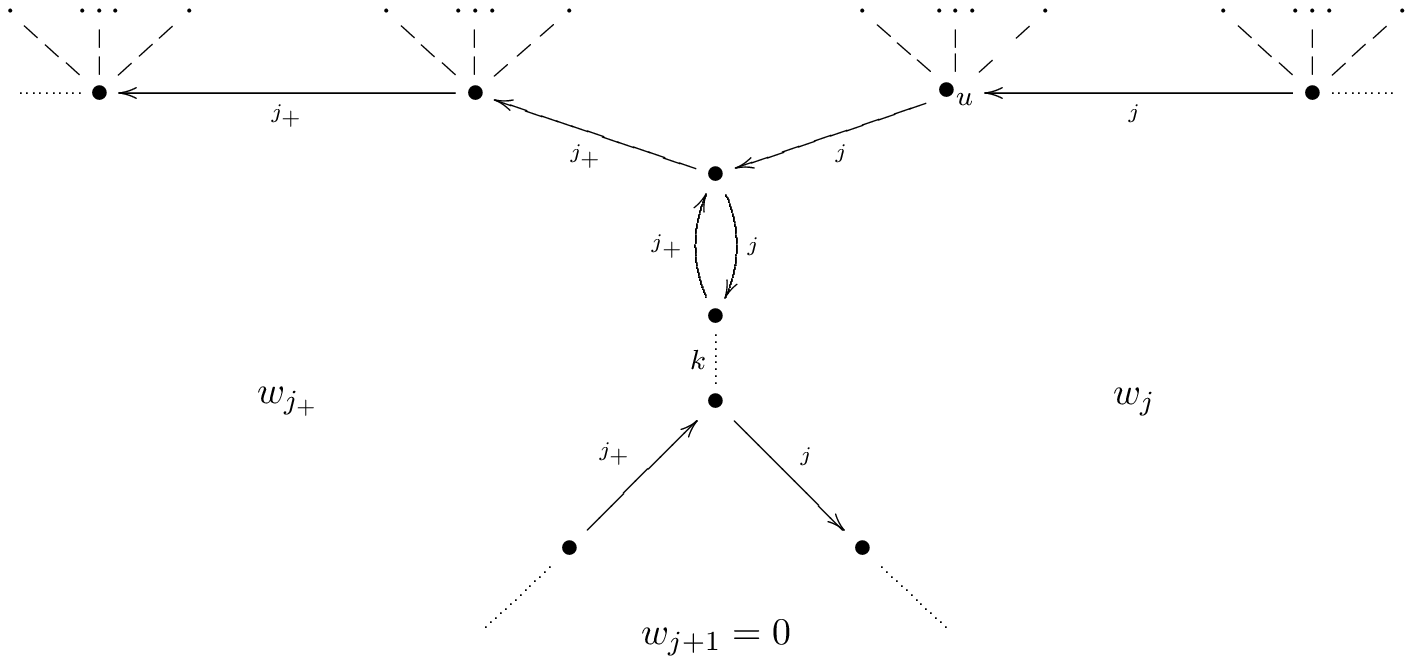}}\\
  \subfloat[(c) Graph $A_j^2(\Gamma)$]{\includegraphics[width=0.8\textwidth]{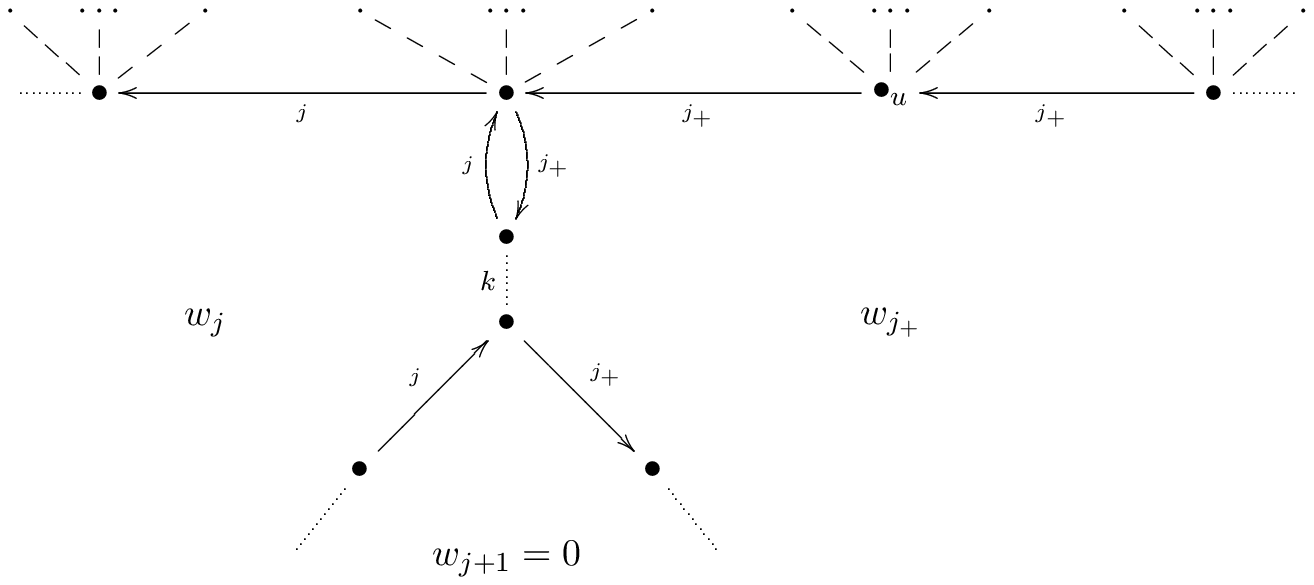}}
  \caption{Case 3, moving a $Y\mhyp$structure.}
  \label{fig:si_action_case2}
\end{figure}
The statement for $\actA_j^{-2}$ is proved similarly.
\end{proof}
Examples of the actions are given in Appendix, Figs.~ \ref{fig:action_example_case1}, \ref{fig:action_example_case3} and \ref{fig:action_example_case2}.

%%%%%%%%%%%%%%%%%%%%%%%%%%%%%%%%%%%%%%%%%%%%%%%%%%%%%%%%%%%%%%%%%%%%%%%%%%%%%%%%

\subsection{Contraction theorems}

\begin{definition}
Let $\Gamma$ be a standard graph and let $\junc_0$ be a junction of $\Gamma.$
The \defin{$\junc_0$-metric} of $\Gamma,$ denoted $|\Gamma|_{\junc_0}$ is defined as
$$|\Gamma|_{\junc_0} = \sum_{\vtex} \left(\deg(\vtex)-2\right)|\vtex - \junc_0|$$
where the sum is taken over all vertices $\vtex$ of $T_\Gamma.$
Here $deg(\vtex)$ is the total degree of the vertex $\vtex$ in $T_\Gamma$ and 
$|\vtex - \junc_0|$ is the length of the shortest path from $\vtex$ to $\junc_0$ in $T_\Gamma.$
(Note that the sum in the right hand side is finite, since only junctions make non-zero contributions.)
\end{definition}

\begin{definition}
A standard graph $\Gamma$ is in \defin{ivy form} if $\Gamma$ is the union of the 
structures connected to a junction $\junc.$
Such junction is called a \defin{root junction}.
\end{definition}

\begin{lemma}
The graph $\Gamma$ is in ivy form if and only if all but one of its junctions are 
$Y\mhyp$junctions.
\end{lemma}
\begin{proof}
This follows from the definitions of the structures.
\end{proof}

\begin{theorem}\label{thm:toIVY}
Let $\Gamma$ be a standard graph.
Then there is a sequence of actions $A^* = \actA_{j_1}^{\pm2},\actA_{j_2}^{\pm2},\dots,$ 
such that $\actA^*(\Gamma)$ is in \defin{ivy form}.
\end{theorem}
\begin{proof}
Assume that $\Gamma$ is not in ivy form.
Let $U$ be the set of junctions in $\Gamma$ that are not $Y\mhyp$junctions. 
Since $\Gamma$ is not in ivy form, $|U|\geq 2$.
Let $\junc_0 \neq \junc_1$ be two junctions in $U$ such that $|\junc_0-\junc_1|$ is maximal.
Let $p$ be the path from $\junc_0$ to $\junc_1$ in $T_\Gamma.$
It is unique since $T_\Gamma$ is a tree.
Let $\vtex$ be the vertex immediately preceeding $\junc_1$ on the path $p.$ 
The edge from $\vtex$ to $\junc_1$ in $T_\Gamma$ is adjacent to 
at least one dominant face with label $j$ such that $w_j \neq 0.$ 
Therefore, there exists a $j\mhyp$edge between $\vtex$ and 
$\junc_1$ in $\Gamma.$
Suppose first that
this $j\mhyp$edge is directed from $\junc_1$ to $\vtex.$
Let us show that in this case $\junc_1$ must be a $j\mhyp$junction, 
i.e., the dominant face labeled $j_+$ is adjacent to $\junc_1$.

Since $\junc_1$ is not a $Y\mhyp$junction, there is a dominant face
adjacent to $\junc_1$ with a label $k\ne j,j_+$.
Hence no vertices of $p$, except possibly $u_1$ may be
adjacent to  $j_+\mhyp$edges.
If $\junc_1$ is not a $j\mhyp$junction, there are no 
$j_+\mhyp$edges adjacent to $\junc_1$.
This implies that any vertex of $\Gamma$ adjacent to a $j_+\mhyp$edge
is further away from $u_0$ that $u_1$.

Let $\junc_2$ be the closest to $\junc_1$ vertex of $\Gamma$
adjacent to a $j_+\mhyp$edge.
Then $\junc_2$ should be a junction of $T_\Gamma$, 
since there are two $j_+\mhyp$edges adjacent to $\junc_2$ in $\Gamma$
and at least one more vertex (on the path from $\junc_1$ to $\junc_2$) 
which is connected to $\junc_2$ by edges with labels other than $j_+$.
Since $\junc_2$ is further away from $\junc_0$ than $\junc_1$
and the path $p$ is maximal, $\junc_2$ must be a $Y\mhyp$junction.
If the $j\mhyp$edges and $j_+\mhyp$edges would meet at $\junc_2$,
$\junc_1$ would be a $j\mhyp$junction.
Otherwise, a subdominant face labeled $j+1$
would be adjacent to both $\junc_1$ and $\junc_2$,
while a subdominant face adjacent to a $Y\mhyp$junction
cannot be adjacent to any other junctions.

Hence $\junc_1$ must be a $j\mhyp$junction.
By Theorem \ref{thm:action}, the action $\actA_{j}^{2}$ 
moves the structure at the $j\mhyp$junction $\junc_1$
one step closer to $\junc_0$ along the path $p,$
decreasing $|\Gamma|_{u_0}$ at least by 1.

The case when the $j\mhyp$edge is directed from $\vtex$ to $\junc_1$
is treated similarly. In that case,
$\junc_1$ must be a $j_-\mhyp$junction, and the action $\actA_{j_-}^{-2}$ moves 
the structure at the $j_-\mhyp$junction $\junc_1$ 
one step closer to $\junc_0$ along the path $p.$

We have proved that if $|U|>1$ then $|\Gamma|_{\junc_0}$ can be reduced. 
Since it is a non-negative integer,
after finitely many steps we must reach a stage where $|U|=1,$ 
hence the graph is in ivy form.
\end{proof}

\begin{remark}
The outcome of the algorithm 
is in general non-unique, and might yield different final 
values of $|A^*(\Gamma)|_{\junc_0}.$
\end{remark}

\begin{lemma}\label{lemma:YVtoVY}
Let $\Gamma$ be a standard graph with a junction $\junc_0$ such that
$\junc_0$ is both a $j_-\mhyp$junction and a $j\mhyp$junction.
Assume that the corresponding structures are of types $Y$ and $V$, in any order. 
Then there is a sequence of actions from the set 
$\{\actA_j^2,\actA_{j_-}^2,\actA_j^{-2},\actA_{j_-}^{-2} \}$ 
that interchanges the $Y\mhyp$structure and the $V\mhyp$structure.
\end{lemma}
\begin{proof}
We may assume that the $Y\mhyp$ and $V\mhyp$structures are attached to 
$\junc_0$ counterclockwise around $\junc_0,$
as in Fig.~\ref{fig:contraction_lemma2}, otherwise we reverse the actions.
By Theorem \ref{thm:action}, the action $\actA_{j}^{2k}$ moves the $V\mhyp$structure $k$ 
steps in the direction of the $j\mhyp$edges.
Choose $k$ so that the $V\mhyp$structure is moved all the way to $\junc_1$, 
as in Fig.~\ref{fig:contraction_lemma3}.
Then $\junc_1$ becomes both a $j_-\mhyp$junction and $j\mhyp$junction, with two $V$-structures attached.
Proceed by applying $\actA_{j_-}^{2k}$ to move the $V\mhyp$structure at 
the $j_-\mhyp$junction $\junc_1$ up to $\junc_0$, as in Fig.~\ref{fig:contraction_lemma1}. 
\begin{figure}
\centering
\includegraphics[width=0.75\textwidth]{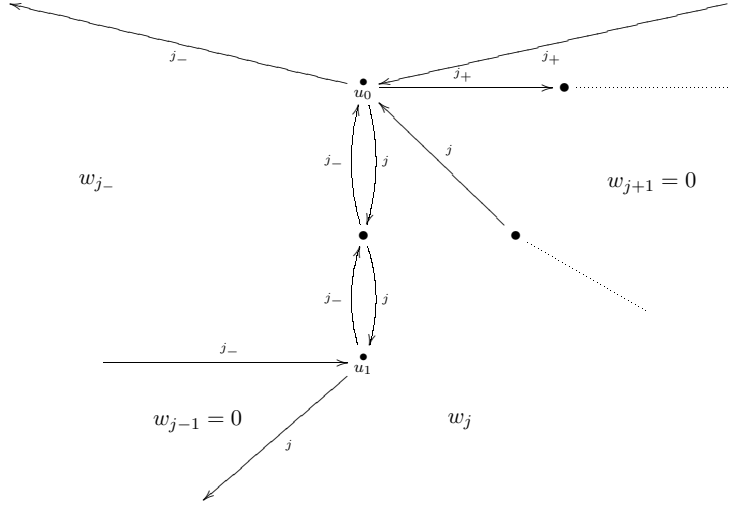}
\caption{Adjacent $Y\mhyp$ and $V\mhyp$structures.}\label{fig:contraction_lemma2}
\end{figure}
\begin{figure}
\centering
\includegraphics[width=0.75\textwidth]{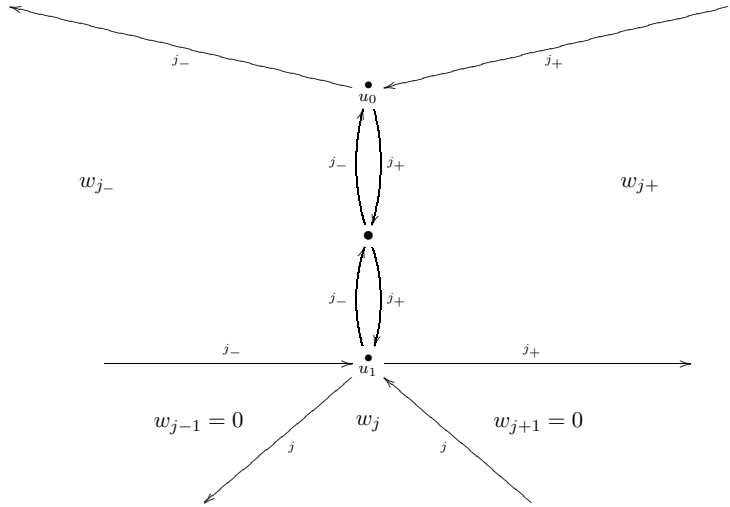}
\caption{Intermediate configuration: two adjacent $V\mhyp$structures.}\label{fig:contraction_lemma3}
\end{figure}
\begin{figure}
\centering
\includegraphics[width=0.75\textwidth]{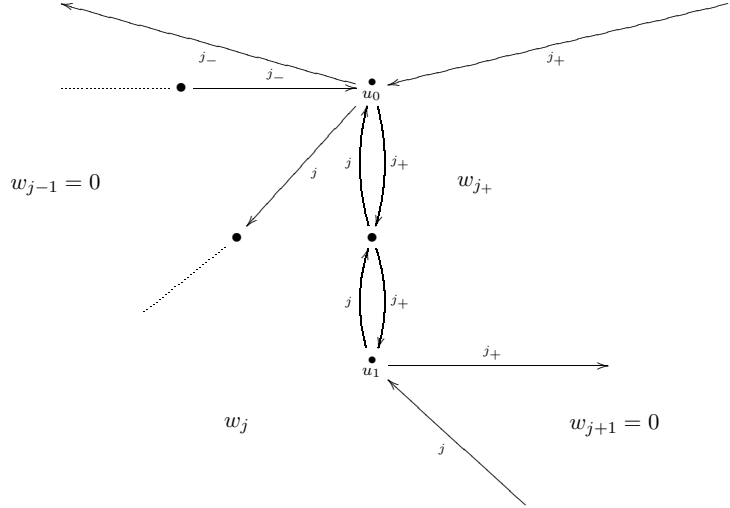}
\caption{$Y\mhyp$ and $V\mhyp$structures exchanged.}\label{fig:contraction_lemma1}
\end{figure}
\end{proof}

\begin{lemma}\label{lemma:IYtoIV}
Let $\Gamma$ be a standard graph with a junction $\junc_0,$
such that $\junc_0$ is both a $j_-\mhyp$junction and a $j\mhyp$junction, 
with the corresponding structures of type $I$ and $Y,$ in any order.
Then there is a sequence of actions from the set 
$\{\actA_j^2,\actA_{j_-}^2,\actA_j^{-2},\actA_{j_-}^{-2} \}$ 
converting the $Y\mhyp$structures to a $V\mhyp$structure.
\end{lemma}
\begin{proof}

We may assume that the $I\mhyp$ and $Y\mhyp$structures are attached to 
$\junc_0$ counterclockwise around $\junc_0,$ as in Fig.~\ref{fig:contraction_trick1},
otherwise, we just reverse the actions.
\begin{figure}
\centering
\includegraphics[width=0.75\textwidth]{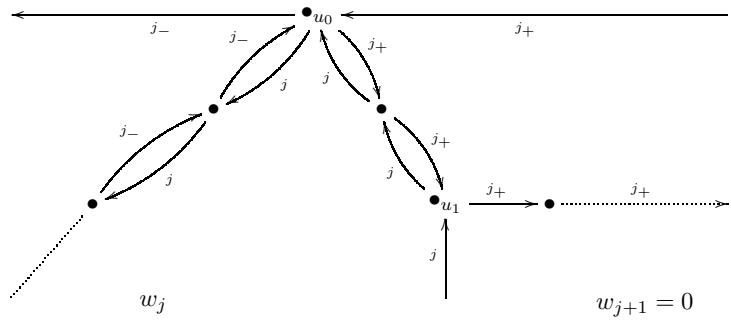}
\caption{Adjacent $I\mhyp$ and $Y\mhyp$structures}\label{fig:contraction_trick1}
\end{figure}
By Theorem \ref{thm:action}, we can apply $\actA_{j_-}^{-2}$ several times to move the 
$I\mhyp$structure down to $\junc_1.$ 
\emph{(For example, in Fig.~\ref{fig:contraction_trick1}, we need to do this twice. 
This gives the configuration shown in Fig.~\ref{fig:contraction_trick2}.)}
\begin{figure}
\centering
\includegraphics[width=0.75\textwidth]{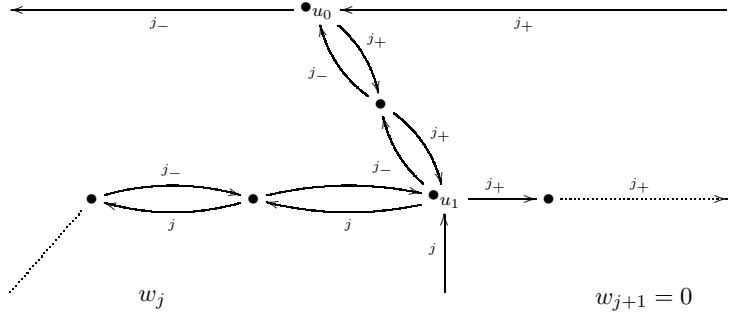}
\caption{Moving the $I\mhyp$structure to $\junc_1$}\label{fig:contraction_trick2}
\end{figure}
Now $\junc_1$ becomes a $j_-\mhyp$junction and a $j\mhyp$structure,
with the $I\mhyp$ and $V\mhyp$structures attached.
Applying $\actA_{j}^{2k},$ we can move the $V\mhyp$structure at $\junc_1$ up to $\junc_0.$
\emph{(In our example, this final configuration is presented in Fig.~\ref{fig:contraction_trick3}.)}
\begin{figure}
\centering
\includegraphics[width=0.75\textwidth]{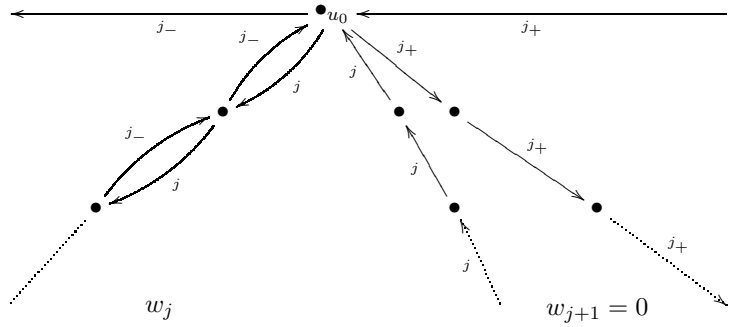}
\caption{Moving the $V\mhyp$structure to $\junc_0$}\label{fig:contraction_trick3}
\end{figure}
Thus the $Y\mhyp$structure has been transformed to a $V\mhyp$structure.
\end{proof}

\begin{theorem}\label{thm:IYtoIV}
Let $\Gamma$ be a standard graph with at least two adjacent dominant faces.
Then there exists a sequence of actions $A^* = \actA_{j_1}^{\pm2} \actA_{j_2}^{\pm2}\dots$ 
such that $A^*(\Gamma)$ have only \emph{one} junction.
\end{theorem}
\begin{proof}
By Theorem \ref{thm:toIVY} we may assume that $\Gamma$ is a graph in ivy form with 
the root junction $\junc_0$. 
The existence of two adjacent dominant faces implies the existence of an $I\mhyp$structure. 
If there are only $I\mhyp$structures and $V\mhyp$structures, then $u_0$ is the only junction of $\Gamma$.
Assume that there is at least one $Y\mhyp$structure.
By Lemma \ref{lemma:YVtoVY}, we may move a $Y\mhyp$structure so that it is counterclockwise 
next to an $I\mhyp$structure. By Lemma \ref{lemma:IYtoIV}, the $Y\mhyp$structure can be transformed to a $V\mhyp$structure, 
and the $Y\mhyp$junction removed. 
This can be repeated, eventually removing all junctions of $\Gamma$ except $\junc_0$.
\end{proof}

\begin{lemma}\label{lemma:YYtoYV}
Let $\Gamma$ be a standard graph with a junction $\junc_0,$
such that $\junc_0$ is both a $j_-\mhyp$junction and a $j\mhyp$junction, 
with two adjacent $Y\mhyp$structures attached.
Then there is a sequence of actions from the set 
$\{\actA_j^2,\actA_{j_-}^2,\actA_j^{-2},\actA_{j_-}^{-2} \}$ 
converting one of the $Y\mhyp$structures to a $V\mhyp$structure.
\end{lemma}
\begin{proof}
This can be proved by the arguments similar to those in the proof of Theorem \ref{thm:IYtoIV}.
\end{proof}

\begin{theorem}\label{thm:toOneY}
Let $\Gamma$ be a standard graph such that no two dominant faces are adjacent.
Then there exists a sequence of actions $A^* = \actA_{j_1}^{\pm2},\actA_{j_2}^{\pm2},\dots,$ 
such that $A^*(\Gamma)$ is in ivy form, with at most one $Y\mhyp$structure.
\end{theorem}
\begin{proof}
One may assume by Theorem \ref{thm:toIVY} that $\Gamma$ is in ivy form, with the root junction $\junc_0$. 
Since no two dominant faces are adjacent, 
there are only $V\mhyp$ and $Y\mhyp$structures attached to $\junc_0$.
If there are at least two $Y\mhyp$structures, we may assume, by Lemma \ref{lemma:YVtoVY},
that two $Y\mhyp$structures are adjacent.
By Lemma \ref{lemma:YYtoYV}, two adjacent $Y\mhyp$structures can be converted to a 
$V\mhyp$structure and a $Y\mhyp$structure.
This can be repeated until at most one $Y\mhyp$structure remains in $\Gamma$.
\end{proof}

\begin{lemma}\label{lemma:boundedfaces}
Let $\Gamma$ be a standard graph such that no two dominant faces are adjacent.
Then the number of bounded faces of $\Gamma$ is finite
and does not change after any action $\actA_j^2$.
\end{lemma}
\begin{proof}
The bounded faces of $\Gamma$ correspond to the edges of $T_\Gamma$
separating two dominant faces.
Since no two dominant faces are adjacent, any two dominant faces
have a finite common boundary in $T_\Gamma$.
Hence the number of bounded faces of $\Gamma$ is finite.
Lemma \ref{lemma:noaction} and Theorem \ref{thm:action}
imply that this number does not change after any action $\actA_j^2$.
\end{proof}

%%%%%%%%%%%%%%%%%%%%%%%%%%%%%%%%%%%%%%%%%%%%%%%%%%%%%%%%%%%%%%%%%%%%%%%%%%%%%%%%

\section{Irreducibility and connectivity of the spectral locus}

In this section, we prove the main results stated in the introduction. 
We start with the following statements.

\begin{lemma}\label{lemma:smooth}
Let $\Sigma$ be the space of all $(\alpha,\lambda)\in\C^d$ 
such that equation (\ref{eq:schroedinger}) admits a solution subdominant
in non-adjacent Stokes sectors $S_{j_1},\dots,S_{j_k},$ $k\leq(d+2)/2.$
Then $\Sigma$ is a smooth complex analytic submanifold of $\C^d$
of the codimension $k-1$.
\end{lemma}
\begin{proof}
Let $f$ be a ratio of two linearly independent solutions of (\ref{eq:schroedinger}),
and let $w=(w_0,\dots,w_{d+1})$ be the set of asymptotic values of $f$
in the Stokes sectors $S_0,\dots,S_{d+1}$.
%TODO: Reformulate 
Then $w$ belongs to the subset $Z$ of ${\Cb}^{d+2}$
where the values $w_j$ in adjacent Stokes sectors are distinct and there are
at least three distinct values among $w_j$.
The group $G$ of fractional-linear transformations of $\Cb$ acts on $Z$ diagonally, 
and the quotient $Z/G$ is a $(d-1)$-dimensional complex manifold.

Theorem 7.2, \cite{bakken} implies that the mapping $W:\C^d\to Z/G$
assigning to $(\alpha,\lambda)$ the equivalence class of $w$ is submersive.
More precisely, $W$ is locally invertible on the subset $\{\alpha_{d-1}=0\}$ of $\C^d$
and constant on the orbits of the group $\C$ acting on $\C^d$ by translations
of the independent variable $z$.
In particular, the preimage $W^{-1}(Y)$ of any smooth submanifold $Y\subset Z/G$
is a smooth submanifold of $\C^d$ of the same codimension as $Y$.

The set $\Sigma$ is the preimage of the set $Y\subset Z/G$
defined by the $k-1$ conditions $w_{j_1}=\dots=w_{j_k}$.
Hence $\Sigma$ is a smooth manifold of codimension $k-1$ in $\C^d$.
\end{proof}

\begin{proposition}\label{thm:connected}
Let $\Sigma$ be the space of all $(\alpha,\lambda)\in\C^d$
such that equation \eqref{eq:schroedinger} admits a solution subdominant
in the non-adjacent Stokes sectors $S_{j_1},\dots,S_{j_k}.$
If at least two remaining Stokes sectors are adjacent,
then $\Sigma$ is an irreducible complex analytic manifold.
\end{proposition}
\begin{proof}
Let $\Sigma_0$ be the intersection of $\Sigma$ with the subspace $\C^{d-1}=\{\alpha_{d-1}=0\} \subset \C^d.$
Then $\Sigma$ has the structure of a product of $\Sigma_0$ and $\C$
induced by translation of the independent variable $z$.
In particular, $\Sigma$ is irreducible if and only if $\Sigma_0$ is irreducible.

Let us choose a point $w=(w_0,\dots,w_{d+1})$ so that
$w_{j_1}=\dots=w_{j_k}=0$, with all other values $w_j$
distinct, non-zero and finite.
Let $\Psi_0$ be a cell decomposition of $\Cb\setminus\{0\}$ defined by the loops $\gamma_j$ starting and ending
at $\infty$ and containing non-zero values $w_j$, as in Section 2.1.

Nevanlinna theory (see \cite{nevanlinnaU,nevanlinnaF}),
implies that, for each standard graph $\Gamma$
with the properties listed in Lemma \ref{lemma:labelingprop},
there exists $(\alpha,\lambda)\in\C^d$ and a meromorphic function $f(z)$ such that
$f$ is the ratio of two linearly independent solutions of (\ref{eq:schroedinger})
with the asymptotic values $w_j$ in the Stokes sectors $S_j$, and $\Gamma$ is the graph
corresponding to the cell decomposition $\Phi_0=f^{-1}(\Psi_0)$.
This function, and the corresponding point $(\alpha,\lambda)$ is defined uniquely up
to translation of the variable $z$.
We can choose $f$ uniquely if we require that $\alpha_{d-1}=0$ in $(\alpha,\lambda)$.
Conditions on the asymptotic values $w_j$ imply then that $(\alpha,\lambda)\in\Sigma'$.
Let $f_\Gamma$ be this uniquely selected function, and $(\alpha_\Gamma,\lambda_\Gamma)$ the
corresponding point of $\Sigma'$.

Let $W:\Sigma'\to Y\subset Z/G$ be as in the proof of Lemma \ref{lemma:smooth}.
Then $\Sigma'$ is an unramified covering of $Y$. 
Its fiber over the equivalence class of $w$ consists of the points
$(\alpha_\Gamma,\lambda_\Gamma)$ for all standard graphs $\Gamma$.
Each action $\actA_j^2$ corresponds to a closed loop in 
$Y$ starting and ending at $w$.
Since for a given list of subdominant sectors
a standard graph with one vertex is unique,
Theorem \ref{thm:IYtoIV} implies that the monodromy action is transitive.
Hence $\Sigma'$ is irreducible as a covering with a
transitive monodromy group (see, e.g., \cite[§5]{khovanskii}).
\end{proof}
This immediately implies Theorem \ref{thm:maintwo}, 
and we may also state the following corollary equivalent to Theorem \ref{thm:mainone}:

\begin{corollary}
For every potential $P_\alpha$ of even degree, 
with $\deg P_\alpha \geq 4$ and with the boundary conditions $y \rightarrow 0$ for $z \rightarrow \pm \infty,$ $z \in \R,$
there is an analytic continuation from any eigenvalue $\lambda_m$ to any other eigenvalue $\lambda_n$ in the $\alpha\mhyp$plane.
\end{corollary}

\begin{proposition}\label{thm:disconnected}
Let $\Sigma$ be the space of all $(\alpha,\lambda)\in\C^d$, for even $d$,
such that equation \eqref{eq:schroedinger} admits a solution subdominant
in the $(d+2)/2$ Stokes sectors $S_0,S_2,\dots,S_d.$
Then irreducible components $\Sigma_k,$ $k=0,\,1,\dots$ of $\Sigma$, which are also its connected
components, are in one-to-one correspondence with the sets of standard
graphs with $k$ bounded faces.
The corresponding solution of \eqref{eq:schroedinger} has $k$ zeros
and can be represented as $Q(z)e^{\phi(z)}$ where $Q$ is a polynomial
of degree $k$ and $\phi$ a polynomial of degree $(d+2)/2$.
\end{proposition}
\begin{proof}
Let us choose $w$ and $\Psi_0$ as in the proof of Proposition \ref{thm:connected}.
Repeating the arguments in the proof of Proposition \ref{thm:connected}, 
we obtain an unramified covering $W:\Sigma'\to Y$ such that its fiber over $w$ 
consists of the points $(\alpha_\Gamma,\lambda_\Gamma)$ for all standard graphs
$\Gamma$ with the properties listed in Lemma \ref{lemma:labelingprop}.
Since we have no adjacent dominant sectors, Theorem \ref{thm:toOneY}
implies that any standard graph $\Gamma$ can be transformed by the monodromy 
action to a graph $\Gamma_0$ in ivy form with at most one $Y$-structure
attached at its $j\mhyp$junction, where $j$ is any index 
such that $S_j$ is a dominant sector. 
Lemma \ref{lemma:boundedfaces} implies that 
$\Gamma$ and $\Gamma_0$ have the same number $k$ of bounded faces.
If $k=0$, the graph $\Gamma_0$ is unique.
If $k>0$, the graph $\Gamma_0$ is completely determined by $k$ and $j$.
Hence for each $k=0,1,\dots$ there is a unique
orbit of the monodromy group action on the fiber of $W$
over $w$ consisting of all standard graphs $\Gamma$ with $k$ bounded faces.
This implies that $\Sigma'$ (and $\Sigma$) has one irreducible component
for each $k$.

Since $\Sigma$ is smooth by Lemma \ref{lemma:smooth}, its irreducible 
components are also its connected components.

Finally, let $f_\Gamma=y/y_1$ where $y$ is a solution of \eqref{eq:schroedinger}
subdominant in the Stokes sectors $S_0,S_2,\dots,S_d$.
Then the zeros of $f$ and $y$ are the same, each such zero belongs to
a bounded domain of $\Gamma$, and each bounded domain of $\Gamma$ contains
a single zero. Hence $y$ has exactly $k$ simple zeros.
Let $Q$ be a polynomial of degree $k$ with the same zeros as $y$.
Then $y/Q$ is an entire function of finite order without zeros,
hence $y/Q=e^\phi$ where $\phi$ is a polynomial.
Since $y/Q$ is subdominant in $(d+2)/2$ sectors, $\deg\phi=(d+2)/2$. 
\end{proof}
The above propisition immediately implies Theorem \ref{thm:mainthree}.

\section{Alternative viewpoint}\label{sec:altview}
In this section, we provide an example of the correspondence between the actions 
on cell decompositions with some subdominant sectors and 
actions on cell decompositions with no subdominant sectors.
This correspondence can be used to simplify calculations with cell decompositions.
We will illustrate our results on a cell decomposition with 6 sectors, the general case follows immediately.

Let $C_6$ be the set of cell decompositions with 6 sectors, none of them subdominant. Let $C_6^{03} \subset C_6$ be the set of cell decompositions such that for any $\Gamma \in C_6^{03},$ the sectors $S_0$ and $S_3$ do not share a common edge in
the associated undirected graph $T_\Gamma.$
Define $D_6^{03}$ to be the set of cell decompositions with 6 sectors where $S_0$ and $S_3$ are subdominant.

\begin{lemma}
There is a bijection between $C_6^{03}$ and $D_6^{03}.$
\end{lemma}
\begin{proof}
Let $\Gamma \in C_6^{03}$ be a cell decomposition, and let $T_\Gamma$ be the associated undirected graph, see section \ref{sec:graphprop}.
Then consider $T_\Gamma$ as the (unique) undirected graph associated with some cell decomposition $\Delta \in D_6^{03}.$
This is possible since the condition that the sectors 0 and 3 do not share a common edge in $\Gamma,$
ensures that the subdominant sectors in $\Delta$ do not share a common edge. Let us denote this map $\pi.$
Conversely, every cell decomposition $\Delta \in D_6^{03}$ is associated with a cell decomposition $\Gamma \in C_6^{03}$
by the inverse procedure $\pi^{-1}.$
\end{proof}

We have previously established that $\sbraid_6$ acts on $C_6$ and that $\sbraid_4$ acts on $D_6^{03}.$ 
Let $\actB_0,\actB_1,\dots,\actB_5$ be the actions generating $\sbraid_6,$ as described in subsection \ref{subsec:actiondef},
and let $\actA_1,\actA_2,\actA_4,\actA_5$ generate $\sbraid_4.$
Let $\sbraid_6^{03} \subset \sbraid_6$ be the subgroup generated by $\actB_1, \actB_2 \actB_{3}  \actB_2^{-1}, \actB_4, \actB_5  \actB_{0} \actB_5^{-1},$ and their inverses. It is easy to see that $\sbraid_6^{03}$ acts on elements in $C_6^{03}$ and preserves this set.

\begin{lemma}
The diagrams in Fig.~\ref{fig:commuting_diagrams} commute.
\begin{figure}
\centering
\includegraphics{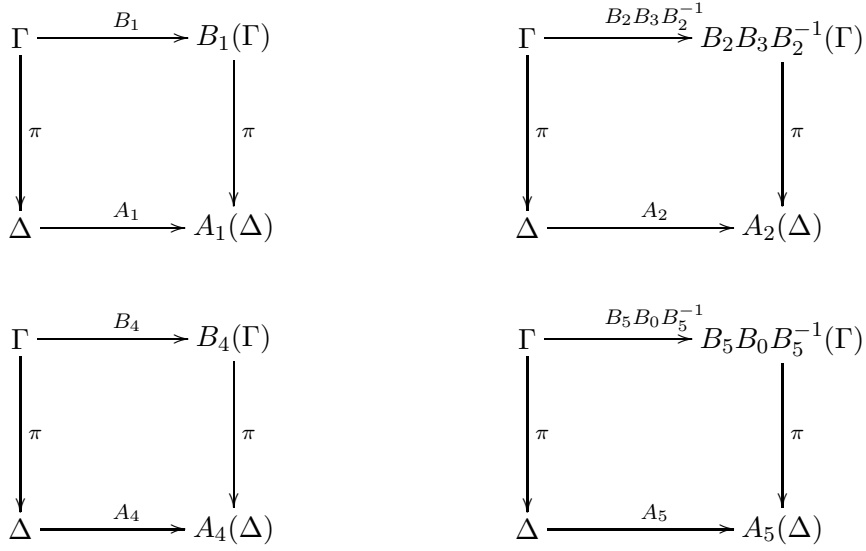}
\caption{The commuting actions}\label{fig:commuting_diagrams}
\end{figure}
\end{lemma}
\begin{proof}
Let $(a,b,c,d,e,f)$ be the 6 loops of a cell decomposition $\Psi_0$ as in Fig.~\ref{fig:curves}, looping around the asymptotic values $(w_0,\dots,w_5).$
Let $\Psi'_0$ be the cell decomposition with the four loops $(b,c,e,f),$ such that if $\Gamma \in C_6^{03}$ is the preimage of $\Psi_0,$
then $\pi(\Gamma)$ is the preimage of $\Psi'_0.$ That is, the preimages of the loops $a$ and $d$ in $\Psi_0$ are removed under $\pi.$

$\actB_j$ acts on $\Psi_0$ and $\actA_j$ acts on $\Psi'_0.$ (See subsection \ref{subsec:actiondef} for the definition.)
We have
\begin{equation}
\begin{gathered}
\actA_1(b,c,e,f)=(bcb^{-1},e,f),\;\actA_4(b,c,d,e)=(b,c,efe^{-1},e).
\end{gathered}
\label{eq:aact14}
\end{equation}
and
\begin{equation}
\begin{split}
\actB_1(a,b,c,d,e,f)&=(a,bcb^{-1},d,e,f),\\
\actB_4(a,b,c,d,e,f)&=(a,b,c,efe^{-1},e,f).
\end{split}
\label{eq:bact14}
\end{equation}
Equation (\ref{eq:aact14}) and (\ref{eq:bact14}) shows that the left diagrams commute,
since applying $\pi$ to the result from (\ref{eq:bact14}) yields (\ref{eq:aact14}).
We also have that
\begin{equation}
\begin{gathered}
\actA_2(b,c,e,f)=(b,cec^{-1},c,f),\actA_5(b,c,e,f)=(f,c,e,fbf^{-1}).
\end{gathered}
\label{eq:aact25}
\end{equation}

We now compute $\actB_3^{-1}\actB_2\actB_3(a,b,c,d,e,f).$ Observe that
we must apply these actions \textit{left to right}:
\begin{equation}
\begin{split}
\actB_3^{-1}\actB_2\actB_3(a,b,c,d,e,f) &= \actB_2\actB_3(a,b,c,e,e^{-1}de,f) \\
&=\actB_3(a,b,cec^{-1},c,e^{-1}de,f) \\
&=(a,b,cec^{-1},c(e^{-1}de)c^{-1},c,f)
\end{split}
\label{eq:bact2}
\end{equation}
A similar calculation gives
\begin{equation}
\actB_0^{-1}\actB_5\actB_0(a,b,c,d,e,f) = (f(b^{-1}ab)f^{-1},f,c,d,e,f,b,f^{-1}),\label{eq:bact5}
\end{equation}
and applying $\pi$ to the results (\ref{eq:bact2}) and (\ref{eq:bact5}) give (\ref{eq:aact25}).
\end{proof}

\begin{remark}
Note that $\actB_j^{-1} \actB_{j-1} \actB_j(\Gamma) = \actB_{j-1} \actB_{j} \actB_{j-1}^{-1}(\Gamma)$
for all $\Gamma \in C_6,$ which follows from basic properties of the braid group.
\end{remark}

The above result can be generalized as follows: Let $C_n$ be the set of cell decompositions with $n$ sectors such that all sectors are dominant.
Let $C_n^{\mathbf{l}} \subset C_n,\;\mathbf{l}=\{l_1,l_2,\dots,l_k\}$ be the set of cell decompositions such that for any $\Gamma \in C_n^{\mathbf{l}},$ no two sectors in the set ${S_{l_1},S_{l_2},\dots,S_{l_k}}$ have a common edge in the associated undirected graph $T_\Gamma.$
Let $D_n^{\mathbf{l}}$ be the set of cell decompositions with $n$ sectors such that the sectors $S_{l_1},S_{l_2},\dots,S_{l_k}$ are subdominant. 
Let $\{\actA_j\}_{j\notin \mathbf{l}}$ be the $n-k$ actions acting on $C_n^{\mathbf{l}}$ indexed as in subsection \ref{subsec:actiondef}.
Let $\{\actB_j\}_{j=0}^{n-1}$ be the actions on $C_n.$
Let $\pi:C_n^{\mathbf{s}} \rightarrow D_n^{\mathbf{s}}$ be the map similar to the bijection above,
where one obtain a cell decomposition in $D_n^{\mathbf{s}}$ by removing edges with a label in $\mathbf{l}$ from a cell decomposition in $C_n^{\mathbf{s}}.$ Then
\begin{equation}
\begin{cases}
\pi(\actB_j(\Gamma)) = A_j(\pi(\Gamma)) &\text{ if } j,j+1\notin \mathbf{l}, \\
\pi(\actB_j^{-1} \actB_{j-1} \actB_j(\Gamma)) = A_j(\pi(\Gamma)),& j\notin \mathbf{l},j+1 \in \mathbf{l}.
\end{cases}
\end{equation}

\begin{remark}
There are some advantages with cell decompositions with no subdominant sectors:
\begin{itemize}
\item An action $\actA_j$ always interchanges the asymptotic values $w_j$ and $w_{j+1}.$
\item Lemma \ref{lemma:labelingprop}, 
item \ref{list:numbering} implies $T_\Gamma$ have no bounded faces iff order of the asymptotic values is a cyclic 
permutation of the standard order.
\end{itemize}
\end{remark}

%%%%%%%%%%%%%%%%%%%%%%%%%%%%%%%%%%%%%%%%%%%%%%%%%%%%%%%%%%%%%%%%%%%%%%%%%%%%%%%%%%%%%%%%5

\section{Appendix}

\subsection{Examples of monodromy action}
Below are some specific examples on how the different actions act on trees and non-trees.

\begin{figure}[ht!]
\centering
\includegraphics[width=0.99\textwidth]{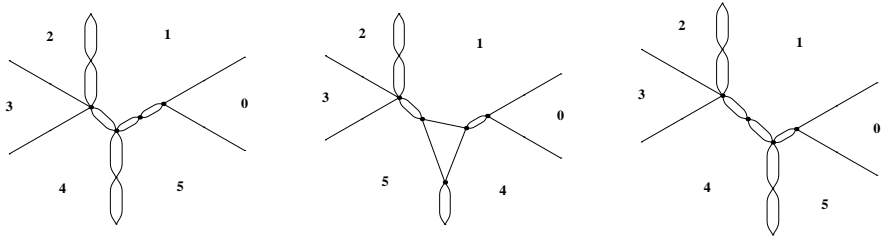}
\caption{Example action of $\actA_4^{-1}$ and $\actA_4^{-2}$ in case 1.}\label{fig:action_example_case1}
\end{figure}

\begin{figure}[ht!]
\centering
\includegraphics[width=0.99\textwidth]{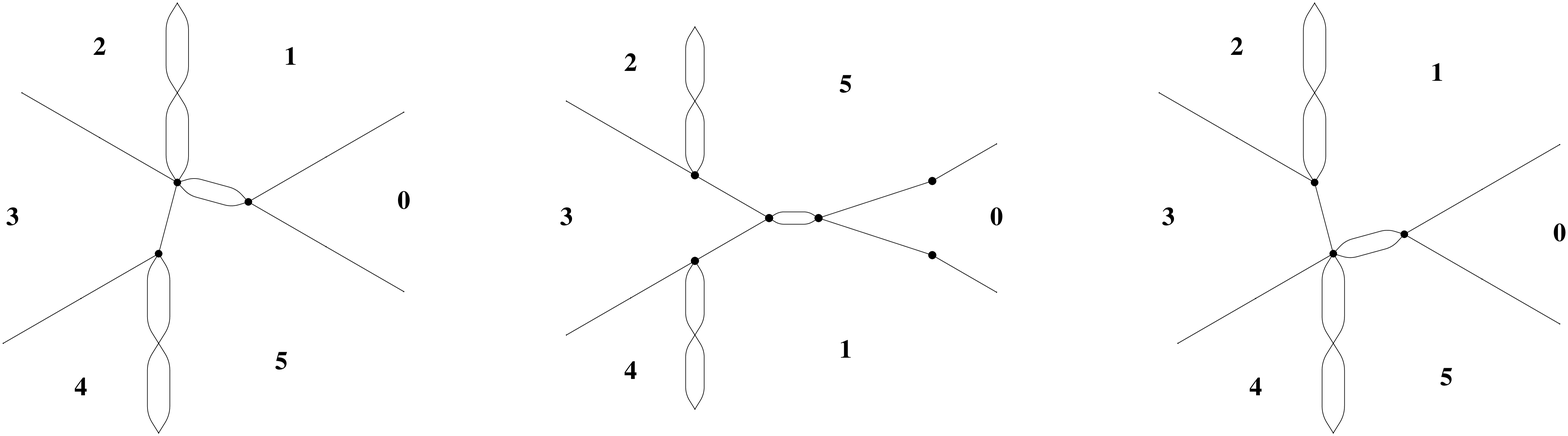}
\caption{Example action of $\actA_5$ and $\actA_5^2$ in case 2.}\label{fig:action_example_case3}
\end{figure}

\clearpage

\bibliographystyle{alpha}
\bibliography{/home/paxinum/public/resources/latex/bibliography} 
\end{document}